\documentclass[aps,superscriptaddress,showpacs,nofootinbib,10pt,notitlepage]{revtex4-2}

\usepackage{graphicx}
\usepackage{amsmath,amssymb}
\usepackage{bbm}
\usepackage[dvipsnames]{xcolor}
\usepackage[utf8]{inputenc}
\usepackage[colorlinks=true,allcolors=BlueViolet,hyperfootnotes=false]{hyperref}
\usepackage{lmodern}
\usepackage[T1]{fontenc}
\usepackage[inline]{enumitem}
\usepackage{accents}
\usepackage{slashed}
\usepackage{orcidlink}
\usepackage{natbib}

\usepackage[normalem]{ulem}

\usepackage{tikz}
\usetikzlibrary{shapes.geometric}

\graphicspath{{Images/}}

\newcommand{\be}{\begin{equation}} \newcommand{\ee}{\end{equation}}
\newcommand{\ba}{\begin{array}{c}} \newcommand{\ea}{\end{array}}
\newcommand{\bea}{\begin{eqnarray}} \newcommand{\eea}{\end{eqnarray}}

\begin{document}

\title{\boldmath Properties of the \texorpdfstring{$T_{cc}(3875)^+$}{Tcc(3875)+} and  \texorpdfstring{$T_{\bar c\bar c}(3875)^-$}{Tcbarcbar(3875)-} (and their heavy-quark spin partners) in nuclear matter}

\newcommand{\ific}{Instituto de F\'{\i}sica Corpuscular (centro mixto CSIC-UV),
Institutos de Investigaci\'on de Paterna,
C/Catedr\'atico Jos\'e Beltr\'an 2, E-46980 Paterna, Valencia, Spain}
\newcommand{\ice}{Institute of Space Sciences (ICE, CSIC), Campus UAB,  Carrer de Can Magrans, 08193 Barcelona, Spain}
\newcommand{\ieec}{Institut d'Estudis Espacials de Catalunya (IEEC), 08034 Barcelona, Spain}
\newcommand{\fias}{Frankfurt Institute for Advanced Studies, Ruth-Moufang-Str. 1, 60438 Frankfurt am Main, Germany}

\author{V.~Montesinos\orcidlink{0000-0002-6186-2777}}
\email{Victor.Montesinos@ific.uv.es}
\author{M.~Albaladejo\orcidlink{0000-0001-7340-9235}}
\email{Miguel.Albaladejo@ific.uv.es}
\author{J.~Nieves\orcidlink{0000-0002-2518-4606}}
\email{jmnieves@ific.uv.es}
\affiliation{\ific}
\author{L.~Tolos\orcidlink{0000-0003-2304-7496}}
\email{tolos@ice.csic.es}
\affiliation{\ice}
\affiliation{\ieec}
\affiliation{\fias}

\definecolor{citecolor}{rgb}{0.15,0.15,0.60}

\begin{abstract}
We discuss the modification of the properties of the tetraquark-like $T_{cc}(3875)^+$ and $T_{\bar c\bar c}(3875)^-$ states in dense nuclear matter. We consider the $T_{cc}^+$ and $T_{\bar c\bar c}^-$ in vacuum as purely isoscalar $D^{\ast} D$ and $\overline{D}{}^{\ast} \overline{D}$ $S$-wave bound states, respectively, dynamically generated from a heavy-quark effective interaction between the charmed mesons. We  compute the $D$, $\overline{D}$, $D^*$, and $\overline{D}{}^{*}$ spectral functions embedded in  a nuclear medium and use them  to determine the corresponding $T_{cc}^+$  and $T_{\bar c\bar c}^-$ self energies and spectral functions. We find important modifications of the $D^{\ast} D$ and $\overline{D}{}^{\ast} \overline{D}$ scattering amplitudes  and of the pole position of these exotic states already for $\rho_0/2$, with $\rho_0$ the normal nuclear density. We also discuss the dependence of these results on the $D^{\ast} D$ ($\overline{D}{}^{\ast} \overline{D}$) molecular component in the $T_{cc}^+$ ($T_{\bar c\bar c}^-$ ) wave-function. Owing to the different nature of the $D^{(*)}N$ and $\overline{D}{}^{(*)}N$ interactions,  we find characteristic changes of the in-medium properties of the $T_{cc}(3875)^+$ and $T_{\bar c\bar c}(3875)^-$,  which become increasingly visible as the density increases. The experimental confirmation of the found distinctive density-pattern will give support to the existence of molecular components in these tetraquark-like states, since in the case they were mostly colorless compact quark structures ($cc\bar \ell\bar \ell$ and $\bar c\bar c \ell \ell$, with $\ell=u,d$), the density behaviors of the $T_{cc}(3875)^+$ and $T_{\bar c\bar c}(3875)^-$ nuclear medium spectral functions, though different, would not likely be the same as those found in this work for molecular scenarios. Finally, we perform similar analyses for the isoscalar $J^P=1^+$ heavy-quark spin symmetry partners of the $T_{cc}^+$ ($T_{cc}^{*+}$) and the $T_{\bar c\bar c}^-$ ($T_{\bar c\bar c}^{*-}$) by considering the $D^{*0}D^{*+}$ and $\overline{D}{}^{*0} D^{*-}$ scattering $T-$matrices.
\end{abstract}

\maketitle

\section{Introduction} \label{sec:intro}

Over the past decades a plethora of new hadronic states has been experimentally observed. More precisely, the spectroscopy of charmonium-like states, the so-called $XYZ$, has received an incredible boost, having the $X(3872)$ \cite{Belle:2003nnu} a prominent and pioneer role. Also the discovery of the $P_c$ and $P_{cs}$ baryonic states by LHCb \cite{LHCb:2015yax,LHCb:2015qvk,LHCb:2019kea,LHCb:2020jpq,LHCb:2022jad}, and more recently mesons, such as $T_{cs}(2900)$ \cite{LHCb:2020bls,LHCb:2020pxc} and $ T_{cc}(3875)^+$ \cite{LHCb:2021vvq,LHCb:2021auc}, have captured the attention of  the hadronic community, as different theoretical interpretations of their nature have been postulated---they can be understood as multiquark states (tetraquarks or pentaquarks), hadroquarkonia states, hadronic molecules, cusps due to kinematic effects, or a mixture of different components (see, for example, the recent reviews \cite{Guo:2017jvc,Brambilla:2019esw,Liu:2019zoy,Dong:2021bvy,Dong:2021rpi,Dong:2021juy,JPAC:2021rxu}).

In particular, the interest on the properties and nature of the $T_{cc}(3875)^+$ state is growing by the day within the hadronic community. This very narrow state is observed in the $D^0D^0 \pi^+$ mass distribution, with a mass of $m_{\rm{thr}} + \delta m_{\rm{exp}}$, being $m_{\text{thr}} = 3875.09\,\text{MeV}$ the $D^{*+} D^0$ threshold and $\delta m_{\rm{exp}} = -360 \pm 40^{+4}_{-0}\,\text{keV}$, and a width $\Gamma = 48 \pm 2^{+0}_{-14}\,\text{keV}$ \cite{LHCb:2021auc}. Among the possible interpretations,  the molecular picture \cite{Dong:2021bvy,Feijoo:2021ppq,Ling:2021bir,Fleming:2021wmk,Ren:2021dsi,Chen:2021cfl,Albaladejo:2021vln,Du:2021zzh,Baru:2021ldu,Santowsky:2021bhy,Deng:2021gnb,Ke:2021rxd,Agaev:2022ast,Meng:2022ozq,Abreu:2022sra,Chen:2022vpo,Albaladejo:2022sux,Dai:2023cyo,Wang:2023ovj} is being supported by its closeness to the $D^0D^{*+}$ and $D^+D^{*0}$ thresholds, whereas the tetraquark interpretation has been put forward \cite{Ballot:1983iv,Zouzou:1986qh}, even before its discovery. However, the proximity of the state to the $D^0D^{*+}$ and $D^+D^{*0}$ thresholds makes it necessary to consider the hadronic degrees of freedom for the analysis of the experimental data \cite{Dong:2021rpi}.

More information on this state in different experimental setups is therefore very welcome in order to further learn about its nature and properties. Recently, the femtoscopic correlation functions for the $D^0D^{*+}$ and $D^+D^{*0}$ channels in heavy-ion collisions (HICs) have become of major interest. Work on that direction has been recently performed for the $T_{cc}(3875)^+$ state in Ref.~\cite{Kamiya:2022thy}, using coordinate space wave functions and potentials, or even more recently in Ref.~\cite{Vidana:2023olz} using momentum space interactions. 

Another possible way to gain some insight about the nature of the $T_{cc}(3875)^+$ is to analyze its behavior under the extreme density and/or temperature conditions present in HICs at RHIC, LHC or FAIR energies. Indeed, analyses of that type have been performed, for example, for the $X(3872)$ state. Using the coalescence model, the ExHIC collaboration \cite{ExHIC:2010gcb,ExHIC:2011say,ExHIC:2017smd} showed that  considering the $X(3872)$ as a molecular state implies a production yield much larger than for the tetraquark configuration, in particular if one also takes into account the evolution in the hadronic phase \cite{Cho:2013rpa,Abreu:2016qci}, due to the fact that the production and absorption cross sections in HICs are expected to be larger for a molecular state. Moreover, the nature of the $X(3872)$ in HICs has been also studied with instantaneous coalescence models \cite{Fontoura:2019opw,Zhang:2020dwn}, a statistical hadronization approach \cite{Andronic:2019wva,Wu:2020zbx}, or using a thermal-rate equation scheme \cite{Wu:2020zbx}. However, these analyses on the production of $X(3872)$ in HICs did not take into account the possible in-medium modification of the $X(3872)$ in the hadronic phase. The inclusion of these modifications has been performed in posterior studies of the $X(3872)$ in a hot meson bath \cite{Cleven:2019cre,Montana:2022inz} and in a dense nuclear medium \cite{Albaladejo:2021cxj}. The in-medium mass shifts of heavy mesons such as the $X(3872)$ and $Z_c(3900)$ have also been studied by means of sum rules \cite{Azizi:2017ubq,Azizi:2020itk}.

In this work we address the behavior of $T_{cc}(3875)^+$ in a nuclear environment, with the
objective of analyzing the finite-density regime of HICs in experiments such as CBM at
FAIR. We follow our previous work on the $X(3872)$ in dense nuclear matter \cite{Albaladejo:2021cxj}. We start from a picture of the $T_{cc}(3875)^+$ generated as a bound state from the leading-order interaction of the $D$ and $D^*$ mesons, constrained by heavy-quark spin symmetry (HQSS). HQSS also allows us to have access to the $D^*D^*$ partner of the $T_{cc}(3875)^+$, which we name the $T^*_{cc}(4016)^+$, and has been predicted by several theoretical groups~\cite{Albaladejo:2021vln,Du:2021zzh,Dai:2021vgf}. We then implement the changes of the $D$ and $D^*$ propagators in nuclear matter in order to obtain the in-medium $T_{cc}(3875)^+$ and $T^*_{cc}(4016)^+$ scattering amplitudes and spectral functions. Later on, we consider generalizations of the $D D^*$ and $D^* D^*$ interactions, allowing for scenarios in which the $T_{cc}(3875)^+$ and $T^*_{cc}(4016)^+$ are not purely molecular states. In this manner, we can extract the modification of the mass and the width of these states in nuclear matter for different scenarios, in view of the forthcoming results on HICs at CBM (FAIR). 

In addition, we also pay attention to the $T_{\bar c \bar c}(3875)^-$ and $T^{*}_{\bar c \bar c}(4016)^-$, antiparticles of the $T_{cc}(3875)^+$ and $T^{*}_{cc}(4016)^+$, and whose properties in vacuum are trivially related to those of  the $T^{(*)+}_{cc}$ by the charge-conjugation symmetry. If these exotic states had a predominant molecular origin, the nuclear environment would induce different modifications to charmed $D^{(*)}D^*$ than to anti-charmed $\overline{D}{}^{(*)}\overline{D}{}^*$ pairs of interacting mesons.  This is due to the different strength of the $D^{(*)}N$ and $\overline{D}{}^{(*)}N$ interactions, which should lead to visible changes among the medium-properties of the $T_{cc}^{(*)+}$ and $T_{\bar c\bar c}^{(*)-}$. These differences become larger as the density increases. The nuclear medium breaks the particle-antiparticle symmetry leading to quite different $D^{(*)}$ and  $\overline{D}{}^{(*)}$ spectral functions. This is similar to what occurs in the strange sector when one studies the properties of kaons and anti-kaons embedded in  dense matter. Kaons ($K^0, K^+$) contain a $\bar s$ antiquark and therefore their strong interaction with nucleons cannot produce hyperons, which however can be excited by $\overline{K}{}^0$ and $K^-$ anti-kaons that provide the negative unit of strangeness (quark $s$) needed to conserve flavor.\footnote{Strangeness measurements exploiting the distinct  $K^0$ and $\overline{K}{}^0$ strong interactions on nucleons have been employed to derive new Bell’s inequalities for entangled $K^0\overline{K}{}^0$ pairs produced in $\phi-$decays~\cite{Bramon:2001tb}. Indeed, if a dense piece of ordinary (nucleonic) matter is inserted along the neutral kaon trajectory, by detecting the products from strangeness conserving strong reactions,  the incoming state is projected either into $K^0$ ($K^0p \to K^+n$) or into $\overline{K}{}^0$ ($\overline{K}{}^0p \to \Lambda \pi^+$, $\overline{K}{}^0n \to \Lambda \pi^0$, $\overline{K}{}^0 n \to p K^-$). Due to the different size of the corresponding cross sections, the slab of nuclear matter acts as a $K^0$ regenerator since the probability of disappearance of the neutral antikaon $\overline{K}{}^0$ is significantly larger.} In the case of  $D^{(*)}N$ interactions, there exists the possibility of exciting the odd-parity spin $J=1/2$ and $3/2$ $\Lambda_c(2595)$ and $\Lambda_c(2625)$ resonances~\cite{Garcia-Recio:2008rjt,Romanets:2012hm}, while in the $\overline{D}{}^{(*)}N$ case, only exotic pentaquarks with negative charm quantum number could be excited~\cite{Gamermann:2010zz}. 

However, if the   $T_{cc}(3875)^+$ and $T_{\bar c\bar c}(3875)^-$ were colorless compact tetraquark structures ($cc\bar \ell\bar \ell$ and $\bar c\bar c \ell \ell$, with $\ell=u,d$), the density behavior of their nuclear medium spectral functions would be presumably different. In this case, the interaction with the medium depends on whether the tetraquark state is composed of two light quarks or two light antiquarks, as they will behave differently in the presence of density. In fact, within the quark model picture, the interaction with the medium would be at the quark level via different color-spin interactions between a $T_{cc}(3875)^+$ and a nucleon or a $T_{\bar c\bar c}(3875)^-$ and a nucleon. Hence, the study of the asymmetrical density-pattern of the properties of the $T_{cc}(3875)^+$ and $T_{\bar c\bar c}(3875)^-$ inside of a nuclear environment could become an interesting additional tool to disentangle the structure (compact or molecular) of the exotic  $T_{cc}(3875)^+$. This is a novel and important result of this work, which did not apply to our previous study  of the  $X(3872)$  in nuclear matter carried out  in Ref.~\cite{Albaladejo:2021cxj},  because the latter state has well defined charge-conjugation.\footnote{Note that the behavior of both $T_{cc}(3875)^+$ and $T_{\bar c\bar c}(3875)^-$ in a hot pion bath will be identical since $D^{(*)}\pi$ and $\overline{D}{}^{(\ast)}\pi$ interactions are equal in the $\text{SU}(2)$ limit.}

The manuscript is organized as follows. In Sec.~\ref{sec:formalism} we present the $D^{\ast} D$ and $\overline{D}{}^{\ast}\overline{D}$ scattering amplitudes and the dynamical generation of the $T_{cc}(3875)^+$, $T_{\bar c\bar c}(3875)^-$ and their heavy-quark spin partners in vacuum and finite density. We start by discussing the $T_{cc}(3875)^+$ and  $T_{\bar c\bar c}(3875)^-$ in the vacuum (Subsec.~\ref{sec:vacuum}) and embedded in isospin-symmetric nuclear matter (Subsec.~\ref{sec:Sigma}). In Subsec.~\ref{sec:medium}, we show the in-medium pseudo-scalar and vector heavy-light meson spectral functions, which determine the density modifications of the $D^{\ast}D$ and $\overline{D}{}^{\ast} \overline{D}$ amplitudes.  We also introduce the $T_{cc}(3875)^+$ and $T_{\bar c\bar c}(3875)^-$ self-energies both in vacuum and in nuclear matter (Subsec.~\ref{ss:TccInMedium}), we discuss the type of interaction kernels to be used in our work (Subsec.~\ref{sec:Tccmolecular}), and we connect to the pole positions in the nuclear medium (Subsec.~\ref{sec:poles}). In Subsec.~\ref{sec:hqss-part} we introduce the heavy-quark spin partners of the $T_{cc}(3875)^+$ and $T_{\bar{c}\bar{c}}(3875)^-$, that is, $T_{c c}^*(4016)^+$ and $T_{\bar c \bar c}^*(4016)^-$. In Sec.~\ref{sec:results} we present our results  for $T_{cc}(3875)^+$ (Subsec.~\ref{ss:resultsTcc}), $T_{\bar c\bar c}(3875)^-$ (Subsec.~\ref{ss:resultsTcbarcbar}), as well as  $T_{c c}^*(4016)^+$ and $T_{\bar c \bar c}^*(4016)^-$ (Subsec.~\ref{ss:resultsheavy-quark}). The conclusions are given in Sec.~\ref{sec:conclusions}.

\section{Formalism} \label{sec:formalism}
In this work we closely follow Ref.~\cite{Albaladejo:2021cxj}, in which the in-medium modifications of $D^{\ast}\overline{D}$ scattering and the $X(3872)$ properties are described. We briefly summarize here this formalism focusing on the appropriate modifications.  

\subsection{\boldmath Vacuum \texorpdfstring{$D^*D$}{D*D(*)} and  \texorpdfstring{$\overline{D}{}^{\ast} \overline{D}$}{barD*barD(*)} scattering amplitudes}
\label{sec:vacuum}
We start by considering the $T_{cc}(3875)^+$ as a $D^*D$ state with isospin and spin-parity quantum numbers $I(J^P) = 0(1^+)$. The $T_{cc}^+$ is thus assumed to be an isoscalar, with a minor isospin breaking from the different masses of the channels involved. This is consistent with the experimental analysis, where no trace of a peak is seen in the partner isospin $I = 1$ channel $D^+D^{*+}$ \cite{LHCb:2021vvq,LHCb:2021auc,Albaladejo:2021vln}.
 We consider the particle basis $\left\{D^{*+}D^0,\, D^{*0}D^+\right\}$ and a heavy-quark effective field theory (HQET) interaction  diagonal in the isospin basis. We only take into account the $S$-wave part of the interaction since the $T_{cc}(3875)^+$ is located almost at the $DD^*$ threshold. In the particle basis, the interaction reads
\begin{equation}
   \mathcal{V} = \frac12
    \begin{pmatrix}
        V_0+V_1 & V_1-V_0 \\
         V_1-V_0 & V_0+V_1 
    \end{pmatrix},
\end{equation}
where $V_0$ and $V_1$ are HQET contact interactions in the isospin $0$ and isospin $1$ channels, respectively. We have used the isospin convention $\bar u
= \vert 1/2,-1/2\rangle$ and  $\bar d
= -\vert 1/2,+1/2\rangle$, which induces $D^0=\vert
1/2,-1/2\rangle$ and $D^+=-\vert
1/2,+1/2\rangle$. The potentials $V_0$ and $V_1$ will be, in general, functions of $s=E^2$, the square of the total energy of the two-meson pair in the center of mass (c.m.) frame.

The unitary $T-$matrix, denoted as $\mathcal{T}(s)$, is obtained by solving the Bethe-Salpeter equation (BSE) in the so-called on-shell approximation~\cite{Nieves:1999bx}:
\begin{equation}
    \mathcal{T}^{-1}(s) = \mathcal{V}^{-1} - \mathcal{G}(s)\,, \label{eq:tmatrix}
\end{equation}
where the diagonal $\mathcal{G}(s)$ matrix is constructed out of the two-meson loop functions,
\begin{equation}
    \mathcal{G}(s) = 
    \begin{pmatrix}
        G_{D^{*+}D^0}(s) & 0 \\
        0 & G_{D^{*0}D^+}(s)
    \end{pmatrix}\,,
\end{equation}
and where
\begin{equation}\label{eq:MesonPropagator}
    G_{U W}(s) = i \int \frac{d^4q}{(2\pi)^4} \Delta_U(P-q) \Delta_W(q), \qquad \Delta_{Y}(q) = \frac{1}{(q^0)^2-\vec{q}^{\,\,2} - m_Y^2 + i\varepsilon} ,
\end{equation}
with $\Delta_Y$ the propagator of a certain $Y$ meson of mass $m_Y$ in the free space\footnote{For simplicity, we neglect the widths of the $D^*$ and $\overline{D}{}^*$ mesons in the vacuum.} and $P^2 = s$. We  will need to introduce an ultraviolet cutoff to regularize the $d^3q$ integration and render the two-point function $G_{U W}$ finite.

The formalism for the $T_{\bar c \bar c}(3875)^-$ state runs in parallel  to that of the $T_{cc}(3875)^+$, making use of invariance under charge-conjugation symmetry in the free-space. Thus, the $\overline{D}{}^*\overline{D}$ unitary $T-$matrix is given also by Eq.\eqref{eq:tmatrix} taking $\left\{ D^{*-}\overline{D}{}^0\,,\overline{D}{}^{*0}D^- \right \}$ now as the particle basis. 

Isospin breaking effects in the unitary $T-$matrices are generated by the kinetic terms in the two meson-loop functions, which disappear when the mass splitting between mesons with different charges are neglected, i.e. $m_{D^{(*)+}}=m_{D^{(*)0}}=m_{D^{(*)-}}= m_{\overline{D}{}^{(\ast)0}}=m_{D^{(*)+}}\equiv m_{D^{(*)}}$. In that limit,  $\mathcal{G}(s)$ becomes a  diagonal matrix. 

\subsection{\boldmath Isoscalar \texorpdfstring{$D^*D$}{D*D(*)} and  \texorpdfstring{$\overline{D}{}^{\ast} \overline{D}$}{barD*barD(*)} scattering amplitudes in isospin–symmetric nuclear matter }
\label{sec:Sigma}

 For simplicity, we will  work here in the isospin limit and  will only consider  the modifications of the $T-$amplitudes due to the changes of the two-particle loop-function  $G_{U W}$ induced by the self-energies $\Pi_Y(q^0,\vec{\, q}\,;\,\rho)$
that the $D^{(*)}$ and $\overline{D}{}^{(*)}$ will acquire as result of their interactions with the nucleons of the medium. The self-energies vanish  in the vacuum ($\rho = 0$), but they produce significant changes in the dispersion relations of  the mesons inside  of  nuclear matter of density $\rho$. 

Indeed, when the mesons are embedded in the nuclear medium, their spectral functions depart from pure delta functions, with the position of the quasi-particle peaks being displaced with respect to the free mass position, and becoming broader as the density increases.  Moreover, richer structures are found produced by several resonant-hole excitations that appear around the quasi-particle peaks~\cite{Tolos:2009nn,Garcia-Recio:2010fiq, Garcia-Recio:2011jcj}.   

The meson spectral functions, $S_{Y=D,\overline{D}, D^*,\overline{D}{}^*}$, are defined through the  K\"allen–Lehmann representation of the propagators, 
\begin{align}\label{eq:dressed-pro}
 \Delta_Y(q\,;\rho) & = \frac{1}{(q^0)^2-\omega_Y^2(\vec{q}^{\,\,2}) - \Pi_Y(q^0,\vec{q}\,;\,\rho)}  = \int_0^\infty d\omega \left( 
 \frac{S_Y(\omega,\lvert \vec{q}\, \rvert)}{q^0 - \omega + i\varepsilon} - 
 \frac{S_{\bar{Y}}(\omega,\lvert \vec{q}\, \rvert)}{q^0 + \omega - i\varepsilon}
 \right)~ 
\end{align}
with $\omega_Y(\vec{q}^{\,\,2})= \sqrt{m_Y^2+\vec{q}^{\,\,2}}$. From the above equation, it follows that for $q^0 >0$
\begin{equation}
    S_{D^{(*)}, \overline{D}{}^{(*)}}(q^0, \vec{\, q}\, ;\,\rho) = -\frac{1}{\pi} \mathrm{Im} \ \Delta_{D^{(*)}, \overline{D}{}^{(*)}}(q^0, \vec{\, q} \,;\,\rho) = -\mathrm{Im} \Pi_{D^{(*)}, \overline{D}{}^{(*)}}(q^0,\vec{q}\,;\,\rho)\frac{\left | \Delta_{D^{(*)}, \overline{D}{}^{(*)}}(q^0,\vec{q}\,;\rho)\right|^2}{\pi} 
\end{equation}
The $S-$wave meson self-energies and the spectral functions can be found, for example, in Ref.~\cite{Albaladejo:2021cxj}. These depend on $q^0$ and the modulus of $\vec q$, but not of any direction when taking the spherical symmetric nuclear medium in the laboratory frame, where it is at rest.\footnote{From now on, we also consider the center of mass of the meson pair system to be at rest in the laboratory frame, and take $\vec{P}=0$, so that hence $P^2 = {P^0}^2 = s$.} In the isospin limit, the isoscalar $D^*D$ [$T(s\,;\,\rho)$] and  $\overline{D}{}^*\overline{D}$ [$\overline{T}(s\,;\,\rho)$] scattering amplitudes inside of the nuclear environment are obtained from the solution of the corresponding single-channel BSE in the on-shell approximation
\begin{subequations}%
\label{eq:TAndTbarmedium}%
\begin{align}
 T^{-1}(s\,;\,\rho) &= V_0^{-1}(s) -\Sigma(s\,;\,\rho) \label{eq:Tmedium} \\
 \overline{T}^{-1}(s\,;\,\rho) &= V_0^{-1}(s) -\overline{\Sigma}(s\,;\,\rho) \label{eq:Tbarmedium}
\end{align}
\end{subequations}
where $\Sigma(s\,;\,\rho)$ and $\overline{\Sigma}(s\,;\,\rho)$ are the density dependent $D^{\ast}D$ ($G_{D^{\ast}D}$) and $\overline{D}{}^{\ast}\overline{D}$ ($G_{\overline{D}{}^{\ast}\overline{D}}$) loop functions, respectively, calculated using Eq.~\eqref{eq:MesonPropagator} with the nuclear dressed meson propagators $\Delta_Y(q\,;\rho)$ introduced in Eq.~\eqref{eq:dressed-pro}. From the spectral representation of the meson propagators, it follows for $E>0$~\cite{Albaladejo:2021cxj}
\begin{subequations}%
\label{eq:SigmaAndSigmabar}
\begin{align}
    \Sigma(s=E^2\,;\,\rho) 
    &= \frac{1}{2\pi^2}\left\{{\cal P} \int_0^\infty d\Omega \left( \frac{f_{D^*D}(\Omega\,;\,\rho)}{E-\Omega+i\varepsilon} - \frac{f_{\overline{D}{}^{\ast}\overline{D}}(\Omega\,;\,\rho)}{E+\Omega-i\varepsilon} \right) - i\pi f_{D^*D}(E\, ;\,\rho) \right\}\label{eq:DstarDLoopfunction} \\
 \overline{\Sigma}(s=E^2\,;\,\rho) 
    &= \frac{1}{2\pi^2}\left\{{\cal P} \int_0^\infty d\Omega \left( \frac{f_{\overline{D}{}^{\ast}\overline{D}}(\Omega\,;\,\rho)}{E-\Omega+i\varepsilon} - \frac{f_{D^*  D}(\Omega\,;\,\rho)}{E+\Omega-i\varepsilon} \right) - i\pi f_{\overline{D}{}^{\ast}\overline{D}}(E\, ;\,\rho) \right\}\label{eq:barDstarDLoopfunction}
\end{align}
\end{subequations}
where ${\cal P}$ stands for the principal value of the integral and in addition
\begin{equation}
    f_{UW}(\Omega\,;\,\rho) = \int_0^\Lambda dq\, q^2 \int_0^\Omega d\omega \ S_U \left( \omega,\, |\vec{\, q}\,|;\,\rho\right) S_W \left(\Omega-\omega,\, |\vec{\, q} \,|;\,\rho\right)\label{eq:deff}.
\end{equation}
In Eq.~\eqref{eq:deff} we have included a sharp cutoff $\Lambda=0.7\,\text{GeV}$ in the integral over momentum to regularize the ultraviolet divergence as we explained in Sec.~\ref{sec:vacuum}. In the free space, the spectral function of charmed $D^{(\ast)}$ and anti-charmed $\overline{D}{}^{(\ast)}$ mesons are equal and reduce to $\delta(q^2-m^2_Y)$.  Hence $\Sigma(s\,;\,\rho=0)=\overline{\Sigma}(s\,;\,\rho=0)$ from which follows that free space masses and widths of the $T_{cc}^+$ and $T_{\bar c\bar c}^-$ are the same, as required by charge-conjugation symmetry. However, in a nuclear environment $S_{D^{\ast}} \ne S_{\overline{D}{}^{\ast}}$, since the charmed and anticharmed meson-nucleon interactions are quite different, as discussed in the Introduction. 

\subsection{Pseudo-scalar and vector heavy-light meson self-energies and spectral functions}
\label{sec:medium}

The meson self-energies are computed following a unitarized self-consistent procedure in coupled channels, as described in Refs.~\cite{Tolos:2009nn,Garcia-Recio:2010fiq} for the $D^{(\ast)}$ mesons and in Ref.~\cite{Garcia-Recio:2011jcj} for the $\overline{D}{}^{(\ast)}$ mesons (see also Ref.~\cite{Tolos:2013gta} for a review). The needed $D^{(\ast)}N$ and $\overline{D}{}^{(*)}N$ $T$--matrices in the free space are obtained by solving a  coupled-channels BSE defined by a  $S$-wave transition  meson-baryon kernel, in the  charm $C=\pm 1$ sectors, derived from an effective Lagrangian that implements HQSS~\cite{Garcia-Recio:2008rjt,Gamermann:2010zz,Romanets:2012hm}. The effective Lagrangian accounts for the lowest-lying pseudoscalar and vector mesons and $1/2^+$ and $3/2^+$ baryons and it reduces to the Weinberg-Tomozawa interaction term in the sector where Goldstone bosons are involved, and it incorporates HQSS in the sector where (anti-)charm quarks participate.  

\begin{figure*}[t]
    \centering
    \includegraphics[width=0.33\textwidth]{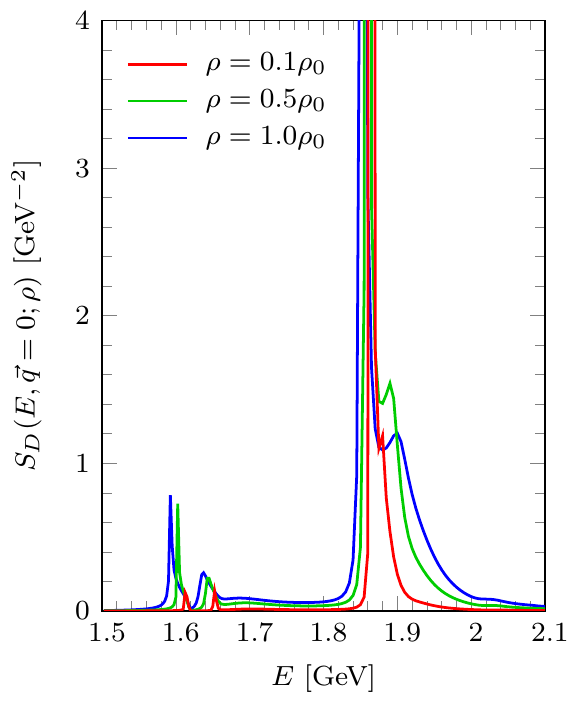}
    \hspace{1cm}
    \includegraphics[width=0.33\textwidth]{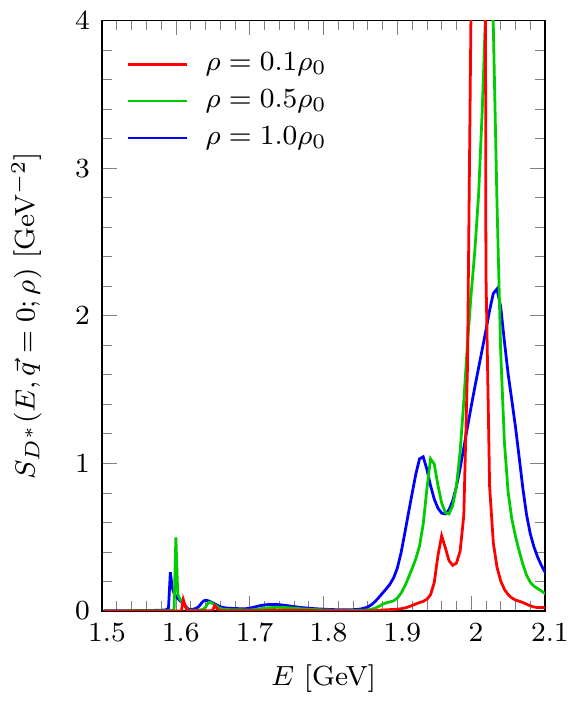}
    \includegraphics[width=0.33\textwidth]{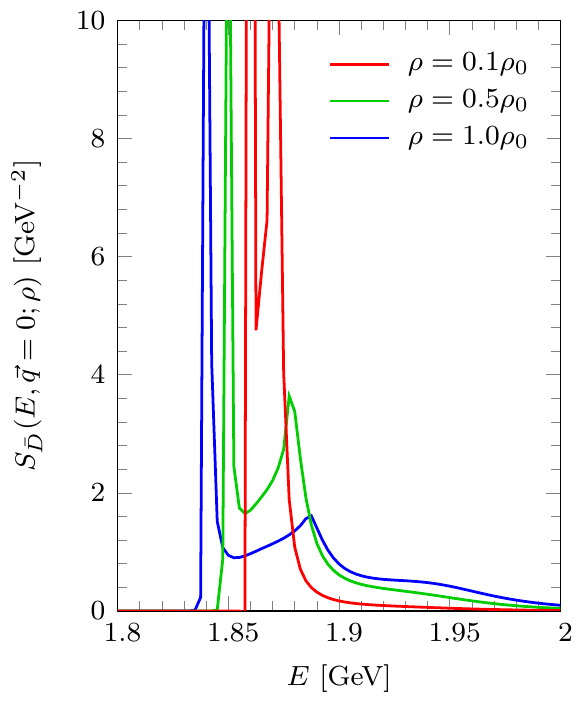}
    \hspace{1cm}
    \includegraphics[width=0.33\textwidth]{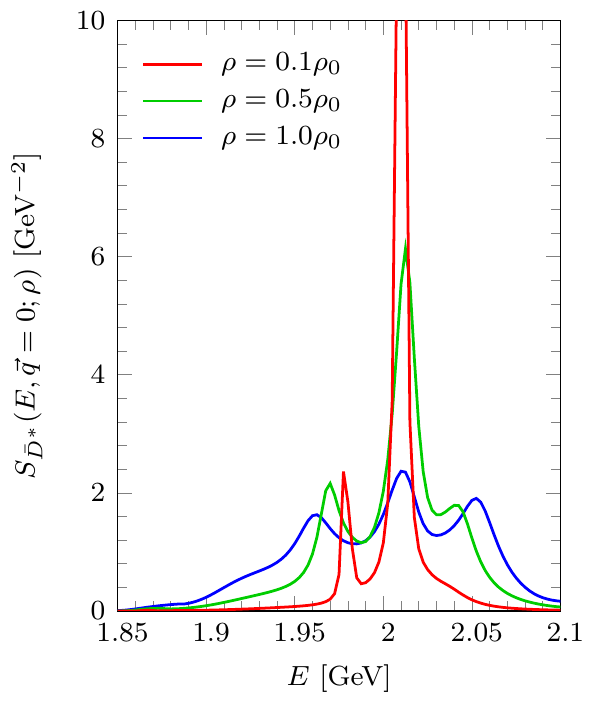}
    \caption{$D$ (upper left-hand side), $D^*$ (upper right-hand side), $\overline{D}$ (lower left-hand side) and $\overline{D}{}^{\ast}$ (lower right-hand side) spectral functions for zero three-momentum ($\vec{\, q} = 0$) as a function of the (anti-)charm meson energy $E$ for three different densities $\rho = 0.1\rho_0$, $0.5\rho_0$ and $\rho_0$.}
    \label{fig:MesonSpectralFunctions}
\end{figure*}

The whole theoretical scheme, both in the vacuum and in the nuclear medium, is briefly summarized in  Section IID of Ref.~\cite{Albaladejo:2021cxj}, where some details can be found. We will only highlight here some of the results found in Refs.~\cite{Tolos:2009nn,Garcia-Recio:2010fiq,Garcia-Recio:2011jcj} for the in-medium $D^{(*)}$ and $\overline{D}{}^{(*)}$ spectral functions. They are plotted in Fig.~\ref{fig:MesonSpectralFunctions} for zero momentum as a function of the (anti-)charmed meson energy $E=q^0$ for three different densities, $\rho/\rho_0 = 0.1$, $0.5$, and $1$, with $\rho_0$ the normal nuclear density ($\rho_0 = 0.17\,\text{fm}^{-3}$). 

The most prominent structure in all cases corresponds to the so called quasi-particle peak, which position ( $q^0=E_\mathrm{qp}$) is obtained from the self-consistent solution of 

\begin{equation}\label{eq:Eqp}
    E^2_\mathrm{qp}(\vec{\, q}\,) = \vec{\, q}^{\,2} + m_{Y= \left\{ D^{(*)}, \overline{D}{}^{(*)} \right\} }^2 + \mathrm{Re} \ \Pi(E_\mathrm{qp}(\vec{\, q}\,),\, \vec{\, q}\,).
\end{equation} 
In addition, these spectral functions show a much richer structure as a result of the presence of several resonance-hole excitations. 

More precisely, the $D$ meson spectral function is depicted in the upper left-hand side panel of Fig.~\ref{fig:MesonSpectralFunctions}. We observe that the $D$ meson quasiparticle peak moves to lower energies with respect to the free mass with increasing density, as already shown in Ref.~\cite{Tolos:2009nn}. Furthermore, several resonance-hole states appear around the quasiparticle peak. On the one hand, the $\Lambda_c(2556)N^{-1}$ and $\Lambda_c(2595)N^{-1}$ excitations appear in the low-energy side of the $D$ spectral function. On the other hand, the $\Sigma_c^* N^{-1}$ state shows up on the right-hand side of the quasiparticle peak. 

As for the $D^*$ meson spectral shown in the upper right-hand side panel, the quasiparticle peak moves to higher energies with density while fully mixing with the sub-threshold $J=3/2$ $\Lambda_c(2941)$ state. The mixing of $J=1/2$ $\Sigma_c(2868)N^{-1}$ and $J=3/2$ $\Sigma_c(2902) N^{-1}$ is seen on the left-hand side tail of the peak. We also observe other dynamically-generated resonance-hole states for lower and higher energies, as described in \cite{Tolos:2009nn}.

With regards to the $\overline{D}$ and $\overline{D}{}^*$ spectral functions, those are shown in the lower left-hand side panel and lower right-hand side one, respectively. The $\overline{D}$ spectral function results from the self-energy of $\overline{D}$, shown in Ref.~\cite{Garcia-Recio:2011jcj}. The quasiparticle peak is located below the $\overline{D}$ mass and also below the $\Theta_c(2805) N^{-1}$ state. The $C=-1$ pentaquark-like $\Theta_c(2805)$ corresponds to a  weakly bound state, seen in the $I=0$, $J=1/2$ amplitude that strongly couples to $\overline{D}N$ and $\overline{D}{}^{\ast}N$, although it has not been detected experimentally yet (see Ref.~\cite{Gamermann:2010zz} for more details).  Also, the upper energy tail of the $\overline{D}$ spectral function shows $I=1$ resonance-hole states.  As for the $\overline{D}{}^*$ spectral function, it depicts the contribution of several $I=0$ and $I=1$ resonant-hole states close to the quasiparticle peak, which is located slightly above to $2\,\text{GeV}$. Those states are described in Ref.~\cite{Gamermann:2010zz}.

\subsection{\boldmath \texorpdfstring{$T_{cc}(3875)^+ \ [T_{\bar c \bar c}(3875)^-]$}{Tcc(3875)+ [Tcc(3875)-]} self-energy in the free space and   in the nuclear medium}\label{ss:TccInMedium}

\begin{figure}[ht]
    \centering
    \includegraphics[height=2cm]{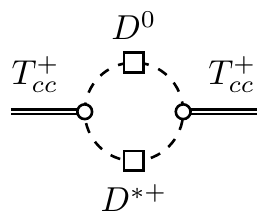}
    \hspace{1cm}
    \includegraphics[height=2cm]{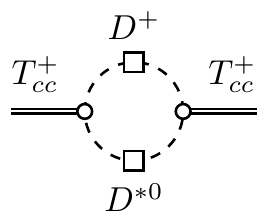}
    \caption{Contributions to the self-energy of the $T_{cc}^+$. The circles represent the bare coupling ($\hat g$) of the $T_{cc}^+$ to the meson pairs, and the squares stand for the interaction of the charm mesons with nuclear matter.}
    \label{fig:TccSelfEnergyDiagrams}
\end{figure}

As in Ref.~\cite{Albaladejo:2021cxj}, let us consider a bare $\widehat{T}_{cc}^+$ field with bare mass $\hat m$ and coupling $\hat g$ to the $D^{\ast}D$ meson pair. We perform the re-summation of the diagrams in Fig.~\ref{fig:TccSelfEnergyDiagrams}, which account for the effects induced by the insertion of internal loops on the $D^{\ast}D$ interaction driven by the  exchange of the bare $\widehat{T}_{cc}^+$ particle. In a first step the bare parameters are renormalized to obtain  the known values of the physical mass ($m_0$) and $D^*D$ coupling ($g_0$) of the $ T_{cc}^+$ in the vacuum. Next, we additionally take into account the renormalization of the heavy-light charmed mesons  inside of the nuclear medium of density $\rho$. The dressed $T_{cc}^+$  propagator is determined by its self-energy, and it reads
\begin{equation}
     \Delta_{T_{cc}^+}(p^2;\, \rho) = \frac{i}{p^2-m_0^2-\Pi_{T_{cc}^+}(p^2;\,\rho)+i\varepsilon}, \qquad  \Pi_{T_{cc}^+}(p^2;\,\rho) = \frac{g_0^2}{1+g_0^2\Sigma_0^\prime(m_0^2)} \left[\Sigma(p^2;\, \rho) - \Sigma_0(m_0^2)\right]. \label{eq:TPropSe}
\end{equation}
The in-medium pole position of the resonance  $m^2(\rho)$ and its density dependent  coupling to the meson pair inside of the nuclear environment are given by~\cite{Albaladejo:2021cxj}
\begin{align}
    m^2(\rho) &=  m_0^2 + \frac{g_0^2}{1+g_0^2\Sigma_0^\prime(m_0^2)} \left[\Sigma[m^2(\rho);\, \rho] - \Sigma_0(m_0^2)\right] , \label{eq:InMediumMassSE}\\
    g^2(\rho) &= \frac{g_0^2}{1-g_0^2\left[\Sigma^\prime[m^2(\rho);\, \rho] - \Sigma_0^\prime(m_0^2)\right]}, \label{eq:InMediumCouplingSE}
\end{align}
where we have defined $\Sigma_0(s) = \Sigma(s;\, \rho=0)$, and the symbol $^\prime$ stands for the derivative with respect $s$. Note that  $m(\rho)$ is in general a complex quantity, with its imaginary part being originated by that of 
$\Sigma[m^2(\rho);\, \rho]$ calculated using Eq.~\eqref{eq:DstarDLoopfunction}. Even assuming that in the free-space the $T_{cc}(3875)^+$ is bound, and therefore $\Sigma_0(m_0^2)$ is real, the in-medium self-energy might acquire
an imaginary part since new many-body decay modes, induced by
the quasielastic interactions of the $D$ and $D^*$ mesons with nucleons, are open. 
The $T_{cc}^+$ spectral function can be evaluated from $S_{T_{cc}^+}(p^2;\,\rho), = -\mathrm{Im} \ \Delta_{T_{cc}^+}(p^2;\, \rho)/\pi$.
The corresponding expressions for the  $T_{\bar c\bar c}^-$ are straightforwardly obtained from those given above by simply 
replacing $\Sigma(s;\, \rho)$ with $\overline{\Sigma}(s;\, \rho)$, calculated using the $\overline{D}$ and $\overline{D}{}^*$ propagators inside of the nuclear medium.

\subsection{\boldmath Isoscalar  \texorpdfstring{$D^*D$}{D*D(*)} and  \texorpdfstring{$\overline{D}{}^*\overline{D}$}{barD*barD(*)} interactions and \texorpdfstring{$T_{cc}(3875)^+ \ [T_{\bar c \bar c}(3875)^-]$}{Tcc(3875)+ [Tcc(3875)-]} molecular contents  in the free space}\label{sec:Tccmolecular}

As we have already mentioned, we work in the isospin limit, and set the in-vacuum masses as $m_{D^{(*)}}=(m_{D^{(*)+}} + m_{D^{(*)0}})/2$. Thus, we cannot consider the physical $T_{cc}(3875)^+$ mass and we will take instead a binding energy of $B = 0.8$ MeV with respect to the $D^*D$ threshold, $m_0 = (m_{D}+m_{D^*}-B)$. This is motivated by the analysis of the $T_{cc}(3875)^+$ as a molecular $D^*D$ state in the isospin limit performed in Ref.~\cite{Albaladejo:2021vln}. To guarantee the existence of pole below threshold at $s=m_0^2$ in the first Riemann sheet of the isoscalar $D^*D$ and $\overline{D}{}^{\ast} \overline{D}$ amplitudes, it follows from Eqs.~\eqref{eq:TAndTbarmedium} that:
\begin{equation}
V^{-1}_0(s=m_0^2) = \Sigma_0(m_0^2) = \Sigma(m_0^2;\, \rho=0) = \overline{\Sigma}_0(m_0^2) = \overline{\Sigma}(m_0^2;\, \rho=0)
\end{equation}
We remind here that a three-momentum sharp cutoff $\Lambda=0.7\,\text{GeV}$ is used to evaluate the two-meson loop function in Subsec.~\ref{sec:Sigma} and hence the numerical value of  $\Sigma_0(m_0^2)$ is completely fixed. For the sake of simplicity, we will drop out from now on the subindex ``0'' in the potential, since we will always refer to the isoscalar amplitudes. 

\begin{figure}[ht]
    \centering
    \includegraphics[height=1.5cm]{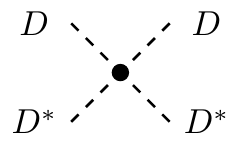}
    \hspace{1cm}
    \includegraphics[height=1.5cm]{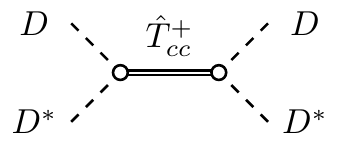}
    \caption{Diagrammatic representation of $V_A$ (left-hand side) and $V_B$ (right-hand side) $D^*D$ potentials.}
    \label{fig:VDiagrams}
\end{figure}

If the potential $V$ was a constant, this is to say does not depend on $s$, then  the $T_{cc}(3875)^+$ and  $T_{\bar c \bar c}(3875)^-$ would be pure $D^{\ast}D$ and $\overline{D}{}^{\ast}\overline{D}$ hadronic-molecules~\cite{Gamermann:2009uq}. As done in the previous analysis on nuclear medium effects of the $X(3872)$~\cite{Albaladejo:2021cxj}, we will consider two families of energy dependent interactions, 
\begin{eqnarray}
V_A(s) &=& \frac{1}{\Sigma_0(m_0^2)} +\frac{\Sigma^\prime_0(m_0^2)}{\left[\Sigma_0(m_0^2)\right]^2}\frac{1-P_0}{P_0}(s-m_0^2),\label{eq:Va} \\
 V_B^{-1}(s) &=& \Sigma_0(m_0^2) -\Sigma^\prime_0(m_0^2)\frac{1-P_0}{P_0}(s^2-m_0^2).\label{eq:Vb}
\end{eqnarray}
where
\begin{equation}
    P_0 = -g_0^2 \Sigma_0^\prime(m_0^2)\,,
\end{equation}
according to the  Weinberg compositeness condition~\cite{Weinberg:1965zz} re-discussed in~\cite{Gamermann:2009uq}, is the molecular probability content of the $D^*D$ bound state of mass $m_0$, and  $g_0^2$ is the residue of the vacuum $T-$matrix [$T(s\,;\,\rho=0)$] at the pole $s=m_0^2$. These interactions correspond to retain the first two orders of the Taylor expansion around of $s=m_0$ either of the potential $V(s)$ (type $A$) or of the inverse of the potential $V^{-1}(s)$ (type $B$). Moreover, it can be shown~\cite{Albaladejo:2021cxj} that $V_B(s)=\hat g^2/(s- \hat m^2)$, and hence this interaction between the $D^*D$ mesons is generated by the exchange of the bare $\widehat{T}_{cc}^+$ introduced in the previous section. The two types of kernels  are diagrammatically represented in Fig.~\ref{fig:VDiagrams}. The $V_A(s)$ potential (left panel of Fig.~\ref{fig:VDiagrams}) depends also on energy and thus it might contain also some contributions related to the exchange of genuine compact quark-model structures, beyond the constant terms which give would rise to purely molecular states~\cite{Gamermann:2009uq}.

\subsection{\boldmath Pole positions of the isoscalar  \texorpdfstring{$D^*D$}{D*D(*)} and  \texorpdfstring{$\overline{D}{}^{\ast} \overline{D}$}{barD*barD(*)} amplitudes in the nuclear medium}
\label{sec:poles}

One could also define the in-medium renormalized pole position and coupling of the $T_{cc}^+$ to the $D^*D$ meson pair from the solution of the BSE of Eq.~\eqref{eq:Tmedium} in nuclear matter using the kernel potentials  $A$ or $B$
\begin{eqnarray}
    0= T^{-1}_{A,B}[m^2(\rho)\,;\,\rho]&=& V^{-1}_{A,B}[m^2(\rho)]-\Sigma[m^2(\rho)\,;\,\rho], \label{eq:mrhoAB} \\
    \frac{1}{g^2(\rho)}&=& \frac{dT^{-1}_{A,B}(s\,;\rho)}{ds}\Big |_{s=m^2(\rho)}
\end{eqnarray}
In the case of the $V_B$ potential  (right panel of Fig.~\ref{fig:VDiagrams}), the above equations lead exactly to Eqs.~\eqref{eq:InMediumMassSE} and \eqref{eq:InMediumCouplingSE} obtained after dressing in the dense medium  the $D^*D$ interaction driven by the exchange of a bare $\hat T_{cc}^+$. However, for the type $A$ interaction, there appear some further density corrections~\cite{Albaladejo:2021cxj} governed by the factor $\xi(\rho)= \Sigma_0(m_0^2)/\Sigma[m^2(\rho);\, \rho]$:
\begin{align}
    m^2(\rho) &=  m_0^2 + \frac{ g_0^2}{1+g_0^2\Sigma_0^\prime(m_0^2)} \left[\Sigma[m^2(\rho);\, \rho] - \Sigma_0(m_0^2)\right]\xi(\rho) , \label{eq:InMediumMassSEVA}\\
    g^2(\rho) &= \frac{g_0^2\xi^2(\rho)}{1-g_0^2\left[\Sigma^\prime[m^2(\rho);\, \rho]\xi^2(\rho) - \Sigma_0^\prime(m_0^2)\right]}. \label{eq:InMediumCouplingSEVA}
\end{align}
We should bear in mind that the $V_A$ potential  contains additional physics to the exchange of a bare $\widehat{T}_{cc}^+$ state, as it is the case for $V_B$ and therefore it should not be surprising that the in-medium $D^*D$ $T-$matrix is not completely determined by the $T_{cc}^+$ self-energy.

In order to obtain the $T_{cc}(3875)^+ $ pole position [$m(\rho)$] when it is produced in a nuclear environment we could either make use of Eqs.~\eqref{eq:InMediumMassSE} and \eqref{eq:InMediumMassSEVA} or we could perform an analytic continuation of the $T-$matrix obtained by solving the BSE (Eq.~\eqref{eq:Tmedium} with interactions of type $A$ or $B$) and search for a pole on the complex plane. In this work we choose the latter of the options and we look for poles of the $T-$matrix in the complex plane. However, independently of the chosen method we face the problem that we need to evaluate the in-medium loop function $\Sigma(s\,;\,\rho)$ for complex values of $s$, and the formula given in Eq.~\eqref{eq:DstarDLoopfunction} is only valid in the real axis. Using it in the complex plane would require knowing the meson spectral functions $S_{U}$ for complex values of its arguments, which cannot be computed within the standard scheme  presented above in Subsec.~\ref{sec:medium}. We follow here Ref.~\cite{Albaladejo:2021cxj}, and we approximate the in-medium loop function as the vacuum two-meson one, but evaluated with complex meson masses. For the case of the $D^*D$ loop function we would write:
\begin{equation}
    \Sigma^{}(E;\,\rho) \simeq G\left[E;\, m_{D^*}^\mathrm{(eff)}(\rho),\,m_{D^{}}^\mathrm{(eff)}(\rho)\right] \equiv G^\mathrm{(eff)}\left(E;\, \rho\right)\,.\label{eq:Sigapprox}
\end{equation}
Even though we treat this as an approximation, the $G^\mathrm{(eff)}$ effective loop function with complex meson masses should replicate the numeric calculation for the $\Sigma$, given that the in-medium modifications for the $\Sigma$ loop function come from the fact that the mesons develop a given width when embedded in the medium. 

The discussion in this subsection can be completely carried over to the $T_{\bar c\bar c}^-$ case, replacing $\Sigma[s;\, \rho]$ with $\overline{\Sigma}[s;\, \rho]$.

\subsection{\boldmath The HQSS partner of the \texorpdfstring{$T_{cc}(3875)^+ $}{Tcc(3875)+}}
\label{sec:hqss-part}

It is a known result that HQSS predicts the existence of degenerate doublets of states.  For the case of the $T_{cc}(3875)^+$, its HQSS partner, which we will name $T_{cc}^{*}(4016)^+$, would be a $I(J^{P}) = 0(1^+)$ state near the $D^*D^*$ production threshold~\cite{Albaladejo:2021vln,Du:2021zzh,Dai:2021vgf}. The formalism developed above for the $T_{cc}(3875)^+$ is easily adapted to describe its HQSS sibling, which will show up as a pole in the isoscalar $D^*D^*$ channel. We will assume that the form of the new potential $V_{\ast}$ is equal to the one used to describe the $T_{cc}(3875)^+$ [Eqs.~\eqref{eq:Va} and \eqref{eq:Vb}], which should be correct up to order $\Lambda_\mathrm{QCD}/m_c$, and we will only change the $T_{cc}^{\ast}(4016)^+$ vacuum mass ($m^*_0$) and the two-meson loop function. The latter is now constructed employing only the nuclear-medium $D^*$ and $\overline{D}{}^{\ast}$ spectral functions, 
\begin{eqnarray}
    \Sigma_*(E;\,\rho) &=& \frac{1}{2\pi^2} \int_0^\infty d\Omega \left( \frac{f_{D^*D^*}(\Omega,\,;\,\rho)}{E-\Omega+i\varepsilon} - \frac{f_{\overline{D}{}^{\ast} \overline{D}{}^{\ast}}(\Omega,\,;\,\rho)}{E+\Omega-i\varepsilon} \right),\label{eq:DstarDstarLoopfunction}
\end{eqnarray}
with $f_{D^*D^*}$ and $f_{\overline{D}{}^{\ast}\overline{D}{}^{\ast}}$ defined in Eq.~\eqref{eq:deff}. Given that the interaction potential is the same in both cases, the most notable source of HQSS breaking comes from the fact that $m_{D^*}-m_D \sim m_\pi$. For the illustrating purposes of this work, we will assume that the vacuum mass $m_0^*$ of the $T_{cc}^{*+}$ state will be shifted from the mass  $m_0$ of the $T_{cc}^+$ by a similar amount, $m_0^*-m_0 \sim m_{D^*}-m_D \sim m_\pi$. 

One can similarly compute the in medium $\overline{D}{}^{\ast}\overline{D}{}^{\ast}$ loop function  $\overline{\Sigma}{}_*(E;\,\rho)$. It will deviate from $\Sigma_*(E;\,\rho)$ for finite nuclear densities because of the different interactions of the $D^*$ and $\overline{D}{}^*$ vector mesons with nucleons.

\section{Results} \label{sec:results}

\subsection{\boldmath Results for the \texorpdfstring{$T_{cc}(3875)^+$}{Tcc(3875)+}}\label{ss:resultsTcc}

Let us now discuss the results that we obtain for the $I(J^P)=0(1^+)$ $D^*D$ amplitude in the nuclear medium $|T(E;\, \rho)|^2$ [Eq.~\eqref{eq:Tmedium}]. For the different plots we use the energy $E$ of the $D^*D$ pair in the c.m. frame, with $s=E^2$. In order to do so, first we need to calculate the in-medium modified $D^*D$ loop function $\Sigma(E;\,\rho)$ [Eq.~\eqref{eq:DstarDLoopfunction}]. Actually the $T-$matrix in the medium of  Eq.~\eqref{eq:Tmedium} can be rewritten as  (we recall here that the subindex “0” in the potential has been suppressed since we will always
refer to the isospin zero amplitudes),
\begin{subequations}%
 \label{eq:Vmenosunoeff}%
 \begin{align}
T^{-1}(s\,;\,\rho) & = V_{\rm eff}^{-1}(s\,;\,\rho) -\Sigma(s\,;\,\rho=0)\,,\\
V_{\rm eff}^{-1}(s\,;\,\rho) & = V^{-1}(s)+\delta\Sigma(s\,;\,\rho)\,,
\end{align}
\end{subequations}
where $\delta\Sigma(s\,;\,\rho)=\left[\Sigma(s\,;\,\rho=0)-\Sigma(s\,;\,\rho)\right]$. 

\begin{figure}[ht]
    \centering
    \includegraphics[width=.55\textwidth]{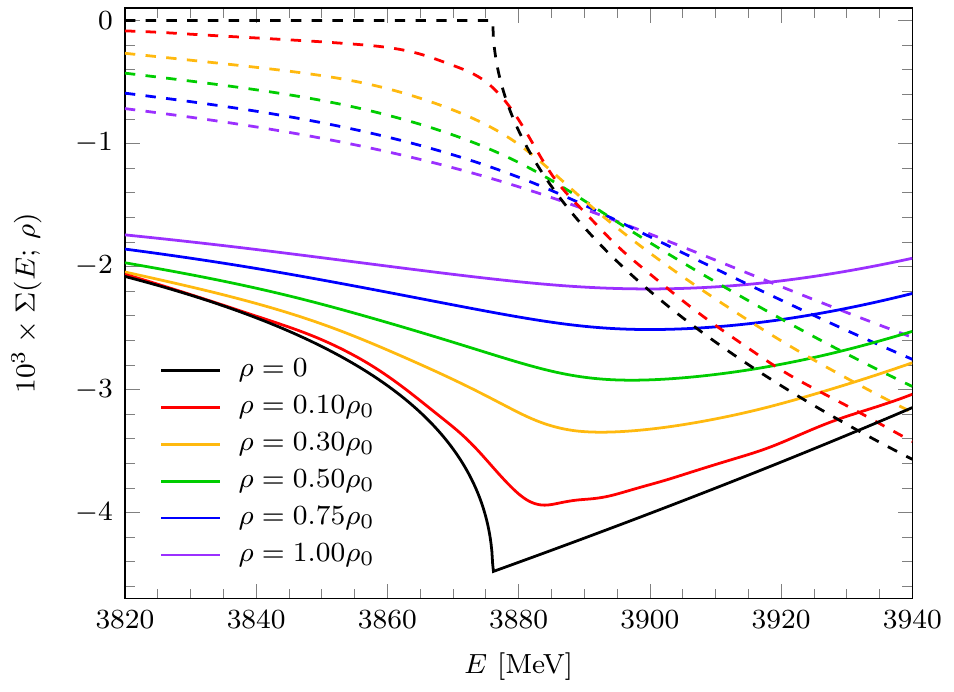}
    \caption{$D^*D$ loop function (Eq.~\eqref{eq:DstarDLoopfunction}) for various values of the nuclear matter density $\rho$ as a function of the $D^*D$ pair energy $E$ in the c.m. frame. The solid and dashed lines stand for the real and imaginary) parts, respectively.}
    \label{fig:DDstarLoop}
\end{figure}

In Fig.~\ref{fig:DDstarLoop} we show $\Sigma(E;\,\rho)$  for different values of the nuclear density $\rho$ ranging from zero to $\rho_0$, where $\rho_0 = 0.17$ fm$^{-3}$ is the normal nuclear matter density. On the one hand, for the imaginary part (dashed lines) we observe that the unitarity cut starting sharply at the $D^*D$ threshold in vacuum gets smoothed out as the density increases. We also observe that the loop function develops an imaginary part even for energies below threshold. This is because the $D$ and $D^*$ mesons acquire some width, given by their spectral functions, when they are embedded in the medium due to the collisions of the $D$ and $D^*$ mesons with nucleons. On the other hand, the real part (solid lines) also flattens for increasing densities, and shifts towards a larger value, ${\rm Re} \, \delta\Sigma(s\,;\,\rho)<0$. This would imply that the effect of the medium is to generate repulsion in the $D^* D$ interaction, when it is attractive in vacuum.  We also note the imaginary part of the self-energy is sizable and comparable to the shift in the real part and  therefore cannot be neglected.

Having calculated the in-medium modified $D^*D$ loop function $\Sigma(E;\,\rho)$, the $D^*D$ $T$-matrix in the nuclear environment can then be determined from the $T_{cc}(3875)^+$ mass and its $D^*D$ probability ($m_0$ and $P_0$) in vacuum ($\rho=0$). For the present analysis, we compute the in-medium effects that enter into the calculation of the amplitude through the vacuum potentials $V_A(E)$ or $V_B(E)$, Eqs.~\eqref{eq:Va} and \eqref{eq:Vb}, respectively, for different values of the molecular probability $P_0$.

%\newpage
\begin{figure*}[t]
    \centering
    \includegraphics[width=.4\textwidth]{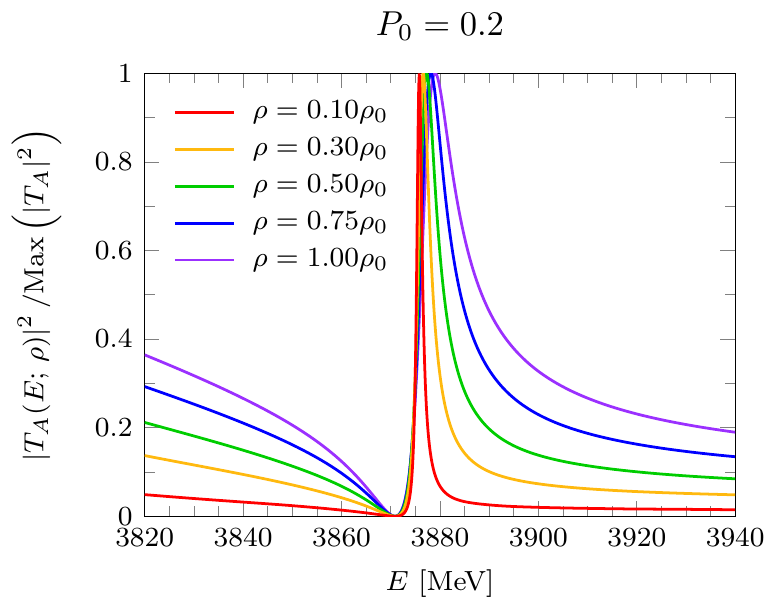}
    \hspace{5mm}
    \includegraphics[width=.4\textwidth]{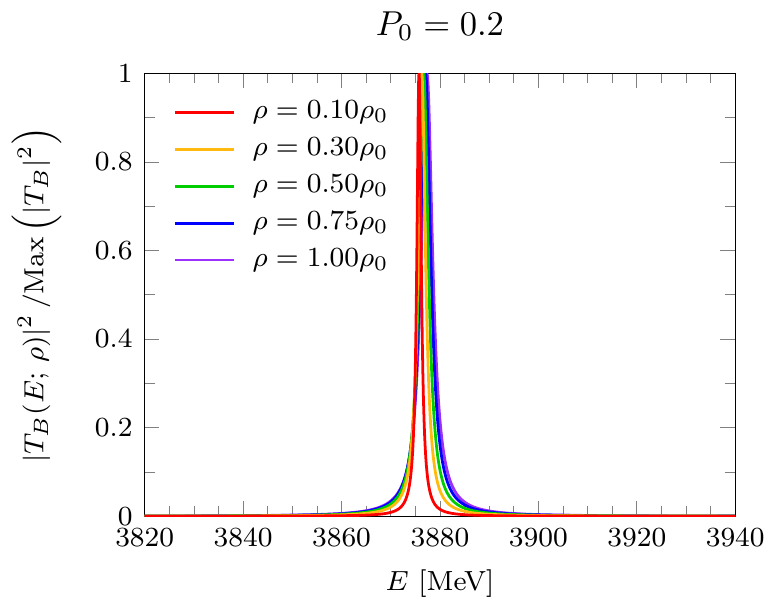}
    \includegraphics[width=.4\textwidth]{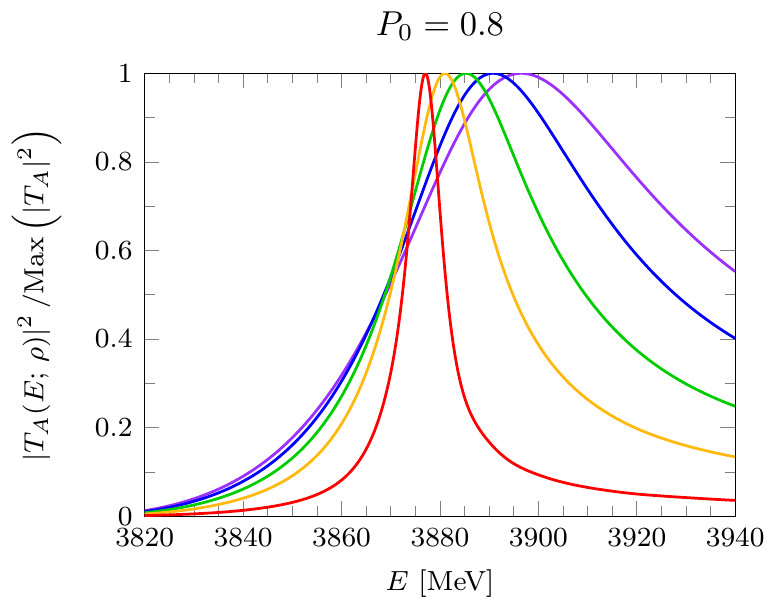}
    \hspace{5mm}
    \includegraphics[width=.4\textwidth]{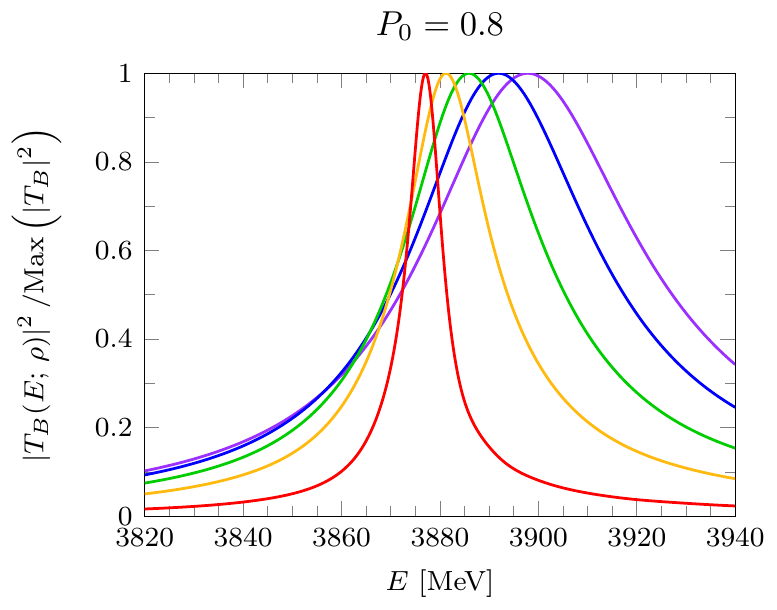}
    \caption{Squared modules of the $D^*D$ amplitudes obtained by solving the BSE using the $V_A(s)$ potential of Eq.~\eqref{eq:Va} (left column) and the $V_B(s)$ potential of Eq.~\eqref{eq:Vb}, as a function of the center-of-mass energy $E$, for different values of the nuclear density $\rho$ (different colors on the graphs) and for different values of the molecular probability $P_0$. Note that the amplitudes have been normalized to be one at their maximum. }
    \label{fig:TABTcc}
\end{figure*}

In Fig.~\ref{fig:TABTcc} we show, for different densities and molecular probabilities $P_0 = 0.2$ and $0.8$, the squared modulus of the amplitudes $T(E;\,\rho)$ normalized to be one at the maximum using the potentials $V_A(s)$ (left column) and $V_B(s)$ (right column). When comparing the amplitudes computed using the $V_A(E)$ potential and the ones obtained from the $V_B(E)$ potential, we conclude that for high values of the molecular $D^*D$ component the predictions of both potentials are very similar. As discussed in Ref.~\cite{Albaladejo:2021cxj}, this results from the fact that the zero of $V_A(s)$ and the bare pole of $V_B (s)$ are far from the energies considered. For small values of $P_0$ ($P_0=0.2$ in the upper plots) both potentials are very different leading to distinct in-medium $T$-matrices, despite giving rise to the same mass ($m_0$) and $D^*D$ coupling ($g_0$) in the free space.

As for the density dependence of the in-medium $T$-matrices at small and large $P_0$, we find that the medium effects on the $T$-matrices are significantly larger for the scenarios where a high molecular probability is considered. For large values of $P_0$, the width increases with density and the maximum peak is shifted to larger energies. This behavior is correlated to the one discussed above for the self-energy in Fig.~\ref{fig:DDstarLoop}. When considering a small molecular component, the changes to the $T_{cc}(3875)^+$ become less important but, as mentioned before, the $T$-matrices differ depending on the potential used. The amplitudes deduced from $V_A(E)$ show the zero that this type of potential has below $E_0$, with the position of the zero being independent of the nuclear density, as discussed in Ref.~\cite{Albaladejo:2021cxj}. However, the amplitude below and above $E_0$ shows a clear dependence on the density as the potential and scattering amplitude vanish.  On the contrary, when using the $V_B(E)$ interaction, we basically observe the peak induced by the bare pole present in the potential. The in-medium effects are in this case even smaller than when considering the $V_A(E)$ potential, and for $P_0=0.2$ the amplitude is almost density independent. Hence, any experimental input on $|T(E;\,\rho)|^2$, in particular for energies below $E_0$, might shed light on the dynamics of the interacting $D^*D$ pair.

\subsection{\boldmath Results for the \texorpdfstring{$T_{\bar c \bar c}(3875)^-$}{Tcc-}}
\label{ss:resultsTcbarcbar}

\begin{figure*}[ht]
    \centering
    \includegraphics[width=.45\textwidth]{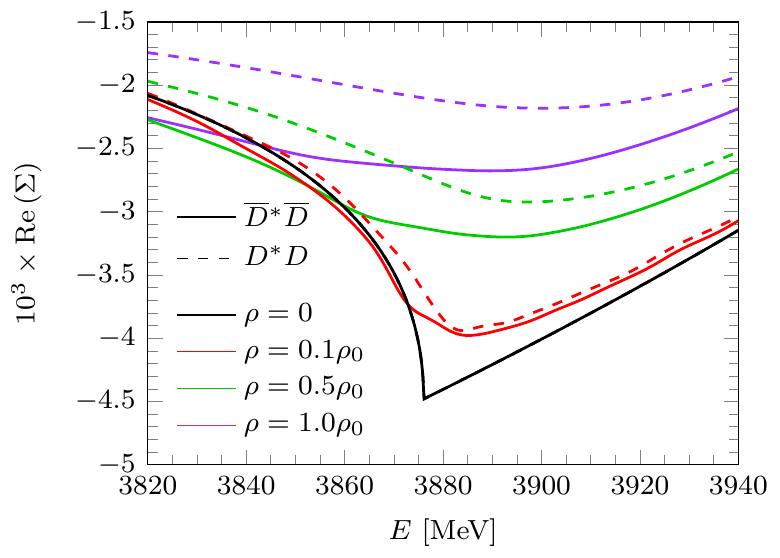}
    \hspace{5mm}
    \includegraphics[width=.45\textwidth]{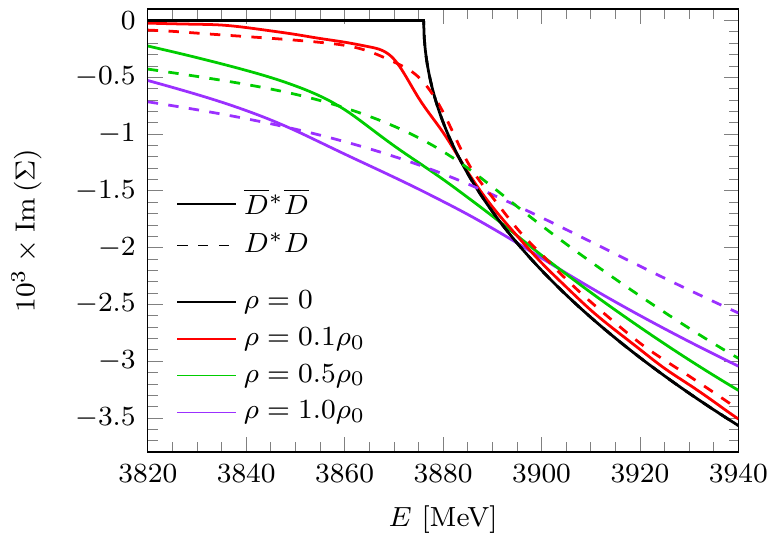}
    \caption{Real (left) and imaginary (right) parts of the $\overline{D}{}^{\ast} \overline{D}$ (solid lines) and $D^*D$ (dashed lines) loop functions of Eqs.~\eqref{eq:barDstarDLoopfunction} and \eqref{eq:DstarDLoopfunction}, respectively. We show results for different values of the nuclear medium density as a function of the c.m. energy of the meson pair.}
    \label{fig:barLoopComparison}
\end{figure*}

\begin{figure*}[t]
    \centering
    \includegraphics[width=.4\textwidth]{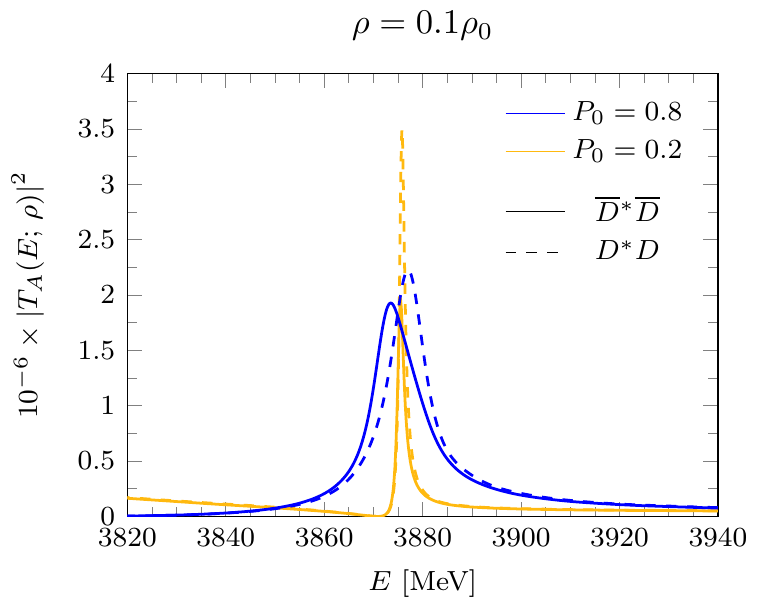}
    \hspace{5mm}
    \includegraphics[width=.4\textwidth]{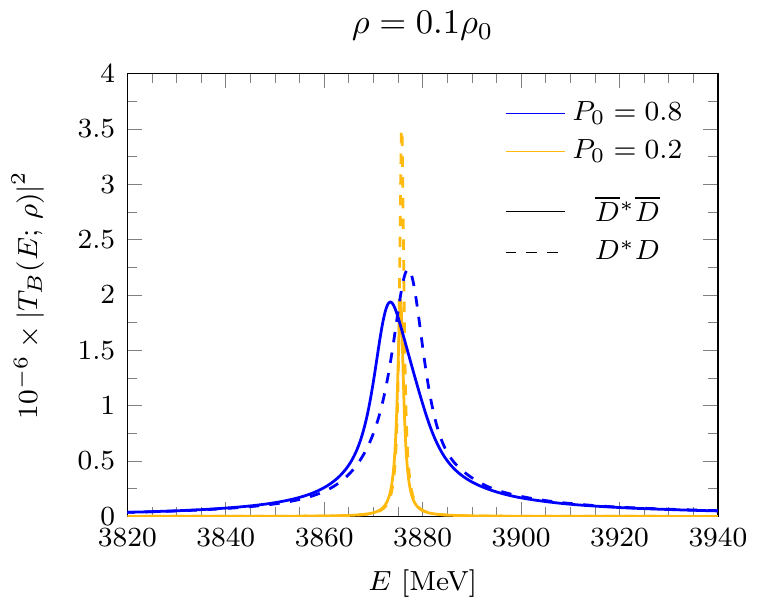}
    \includegraphics[width=.4\textwidth]{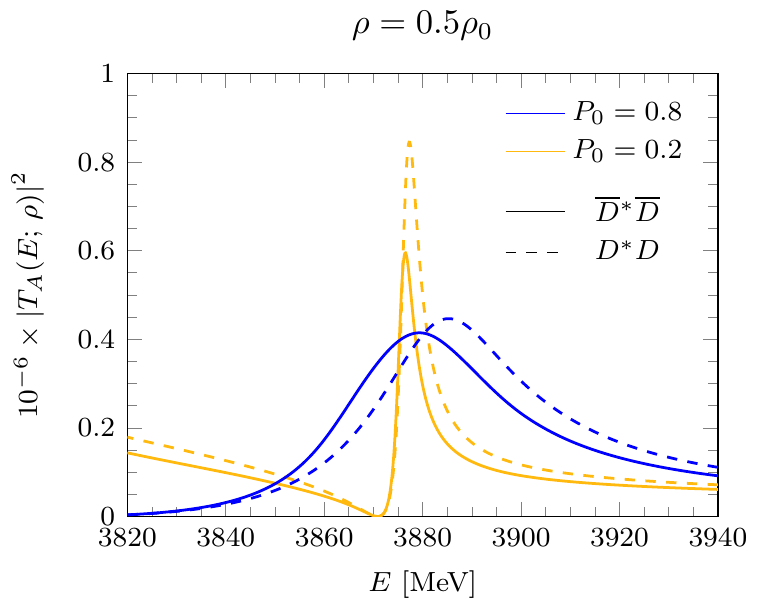}
    \hspace{5mm}
    \includegraphics[width=.4\textwidth]{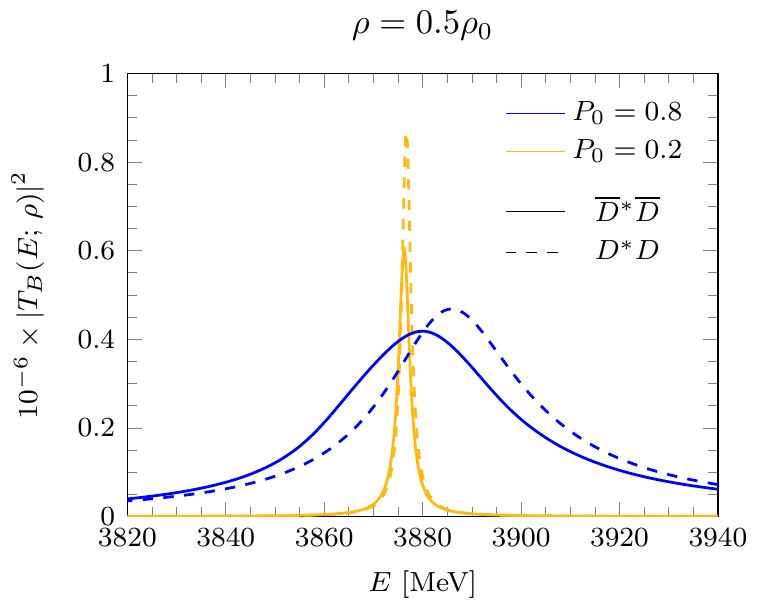}
    \includegraphics[width=.4\textwidth]{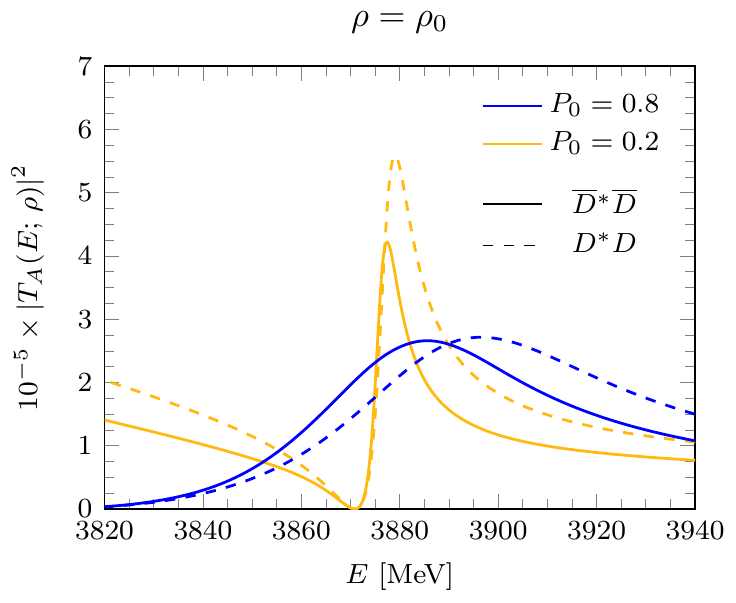}
    \hspace{5mm}
    \includegraphics[width=.4\textwidth]{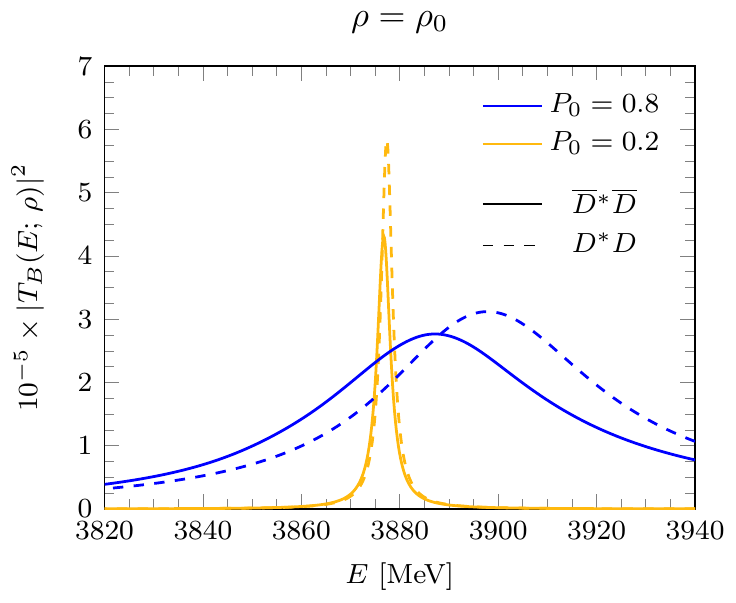}
    \caption{In medium $\overline{D}{}^* \overline{D}$ (solid lines) and $D^* D$ (dashed lines) modulus square amplitudes obtained by solving the BSE using the $V_A(s)$ (left) and $V_B(s)$ (right) potentials, for vacuum molecular probabilities $P_0=0.2$ (orange) and $P_0=0.8$ (blue), and for different nuclear densities $\rho$.}
    \label{fig:TABbarTccP0Comparison}
\end{figure*}

We now turn our attention into comparing the results obtained for the $T_{\bar c \bar c}(3875)^-$ with those presented in Sec.~\ref{ss:resultsTcc} for the $T_{cc}(3875)^+$. We recall here that this is an important novelty with respect to the analysis in Ref.~\cite{Albaladejo:2021cxj} for the $X(3872)$, where this distinction, as explained earlier, does not apply since the $X(3872)$ has well defined $C-$parity. Let us start by discussing Fig.~\ref{fig:barLoopComparison}, where we simultaneously show the energy dependence of the $\overline{D}{}^* \overline{D}$ (solid lines) and the $D^* D$ (dashed lines) loop functions for various nuclear densities.  Both,  real and imaginary parts of the $\overline{D}{}^* \overline{D}$ and $D^* D$ loop functions are the same in vacuum  ($\rho=0$) thanks to charge-conjugation symmetry, which ensures that the $\overline{D}{}^* \overline{D}$ and the $D^* D$ meson pairs have the same masses. However, when considering a density different from zero (even as small as $0.1\,\rho_0$), notable differences appear between both loop functions. This distinctive density-pattern stems from the very different $D^{(*)}N$ and $\overline{D}{}^{(*)}N$ interactions, which were already apparent in the spectral functions presented in Fig.~\ref{fig:MesonSpectralFunctions}. 

We can also define for the in-medium $\overline{D}{}^{\ast}\overline{D}$ pair an effective potential $\overline{V}_{\rm eff}(s\,;\,\rho)$, as done in Eq.~\eqref{eq:Vmenosunoeff} for the $ D^* D $ system, and since the free space terms are equal then it  follows
\begin{equation}
    \overline{V}_{\rm eff}^{-1}(s\,;\,\rho)- V_{\rm eff}^{-1}(s\,;\,\rho) = \frac{V_{\rm eff}(s\,;\,\rho)-\overline{V}_{\rm eff}(s\,;\,\rho)}{V_{\rm eff}(s\,;\,\rho)\overline{V}_{\rm eff}(s\,;\,\rho)}=\Sigma(s\,;\,\rho)-\overline{\Sigma}(s\,;\,\rho) \label{eq:diff-veff}
\end{equation}

Focusing now on the real part of the loop function for different densities shown in Fig.~\ref{fig:barLoopComparison}, we observe that the $\overline{D}{}^* \overline{D}$ real parts always lie below the $D^* D$, with  the difference being more prominent for energies below threshold. The fact that the real part of the $\overline{D}{}^* \overline{D}$ loop function for all densities is smaller than its $D^* D$ counterpart, and both are negative imply that  $\text{Re}\left[\Sigma(s\,;\,\rho)-\overline{\Sigma}(s\,;\,\rho)\right]>0$. This would mean that the medium generates in general a more repulsive interaction in the case of the $T_{cc}^+$ than in the case of the $T_{\bar c \bar c}^-$, as can be deduced from Eq.~\eqref{eq:diff-veff} above. Thus, we might expect to generate the $T_{cc}^+$ at larger energies than the $T_{\bar c \bar c}^-$. As for the imaginary part of the loop function, the one for $T_{\bar c \bar c}^-$ is comparable to the shift in the real part for all densities and should not be neglected, as already seen for the $T_{cc}^+$ in Sec.~\ref{ss:resultsTcc}. We also see that the density-dependent imaginary parts of the $\overline{D}{}^* \overline{D}$ loop change with energy in a more abrupt manner as compared to the ones for the $D^* D$ case. As a consequence, for the smaller energies below the two-meson threshold we find that $|\text{Im} \overline{\Sigma} | < |\mathrm{Im} \Sigma |$, while for energies well above the threshold we have $|\text{Im}  \overline{\Sigma} | > |\mathrm{Im} \Sigma|$. The imaginary parts for $\overline{D}{}^* \overline{D}$ and  $ D^* D$ become comparable for energies which are below but near the vacuum threshold. However, it is not possible to determine whether $T_{\bar c \bar c}^-$ or $T_{cc}^+$ will have a larger width. This is due to the fact that the widths of the states depend on the energy at which they are produced for a given density and we expect the energy to be different.  It could happen that both states have similar widths if they are produced close to the two-meson threshold as the imaginary parts of the two-meson loop functions become alike.
 
Next, in Fig.~\ref{fig:TABbarTccP0Comparison} we show several plots containing the modulus squared of the in medium $\overline{D}{}^* \overline{D}$ and $D^* D$ $T$-matrices (solid and dashed lines, respectively), both computed using the BSE of Eqs.~\eqref{eq:Tbarmedium} and \eqref{eq:Tmedium} as well as using the type-A (left column) and type-B (right column) interaction kernels. We consider three different values for the density (upper rows for $0.1\rho_0$, middle rows for $0.5\rho_0$ and lower rows for $\rho_0$) and the values $P_0=0.2$ (orange lines) and $P_0=0.8$ (blue lines) for the molecular probability. 

We observe that the width of the $T_{\bar c \bar c}^-$ grows with increasing density, being this effect more notable for high values of $P_0$, in a similar manner as for the $T_{cc}^+$ state, as already discussed in Sec.~\ref{ss:resultsTcc}. Differences between the position and the width of the $T_{\bar c \bar c}^-$ and $T_{cc}^+$ states arise with $P_0$ and density. On the one hand, we find that the position of the $T_{\bar c \bar c}^-$ peak always lies below the $T_{cc}^+$ peak when considering high enough values of the molecular probability and density. However, the difference in energy between both states is almost not noticeable for low values of $P_0$ and density, as expected. On the other hand, we observe that the $T_{\bar c \bar c}^-$ state tends to be narrower than the $T_{cc}^+$ for high enough values of the molecular probability and density. However, this effect is not as pronounced as the shift of the peaks, and it is difficult to appreciate in the plots of Fig.~\ref{fig:TABbarTccP0Comparison}. In summary, we can conclude that the behaviors of the $T_{cc}^+$ and $T_{\bar c \bar c}^-$ are quite different when they are embedded in a nuclear medium, and they are very sensitive to their molecular probability in the free space. 

\begin{figure*}[t]
    \centering
    \includegraphics[height=.4\textwidth]{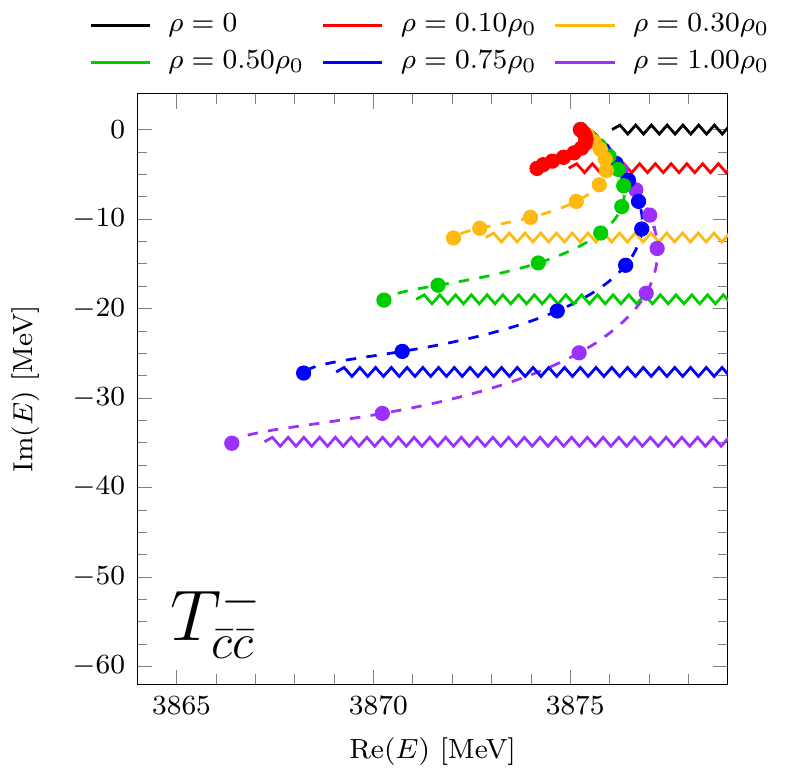}
    \hspace{5mm}
    \includegraphics[height=.4\textwidth]{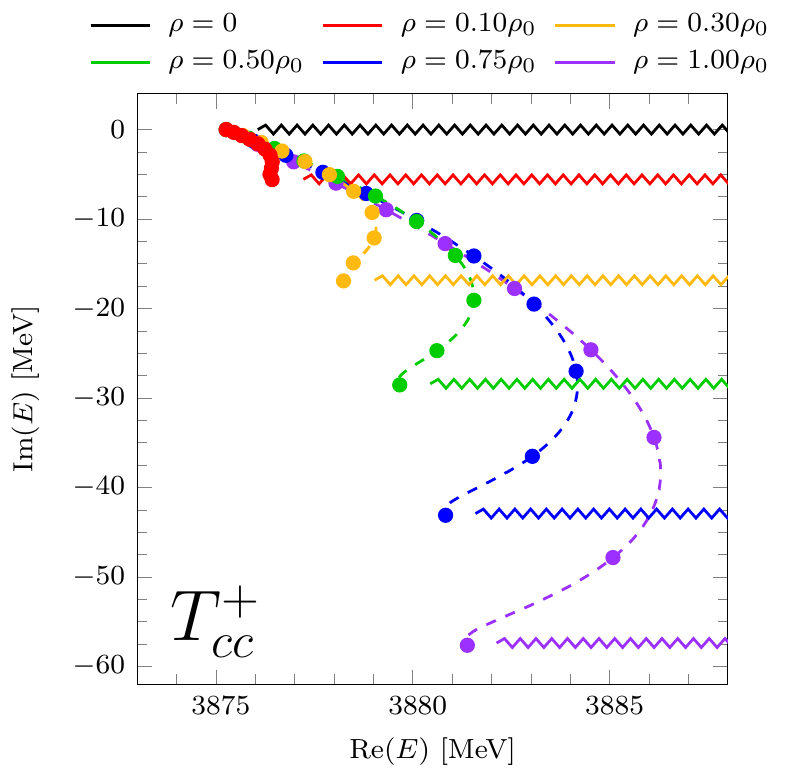}\\
    \vspace{0.5cm}
    \includegraphics[height=.4\textwidth]{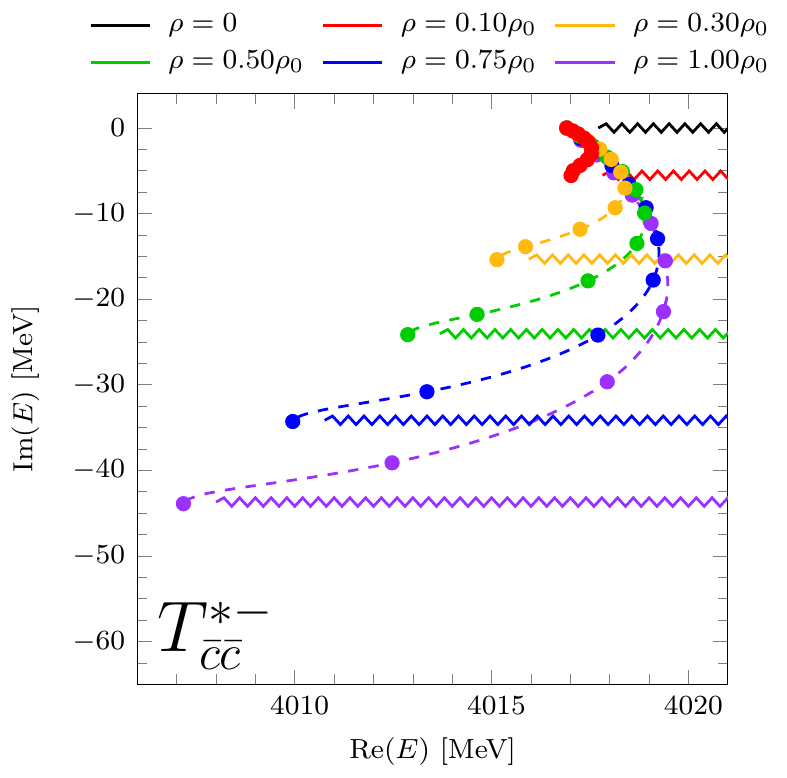}
    \hspace{5mm} 
    \includegraphics[height=.4\textwidth]{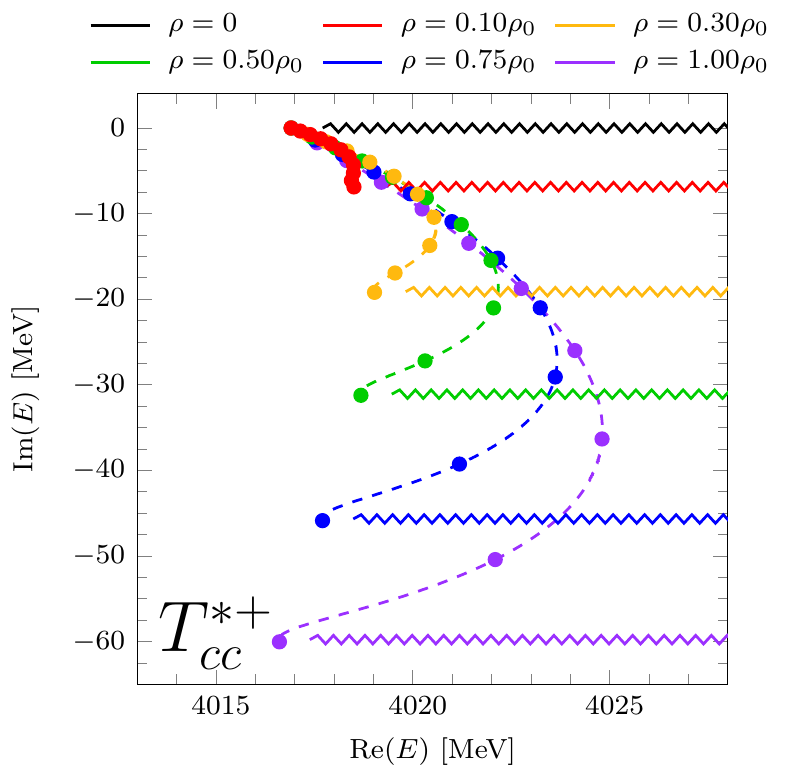}
    \caption{Top: Complex pole positions of the $T_{\bar c \bar c}(3875)^-$ (left) and the $T_{cc}(3875)^+$ (right) for different values of the density ($\rho$)  and vacuum molecular probabilities ($P_0$) obtained using the potential $V_A(s)$. The points that lie on the dashed lines correspond to results for different values of $P_0$, which vary from 0 (right upper end) to 1 (left lower end) in steps $\Delta P_0=0.1$. The zigzag lines represent the cut of the effective loop function $G^{\mathrm{(eff)}}(s;\, \rho)$ for different densities, as detailed in Sect.IIIB of Ref.~\cite{Albaladejo:2021cxj}.
    Bottom: Same as the top plots, but for the $T_{\bar c \bar c}^*(4016)^-$ (left) and the $T_{cc}^*(4016)^+$, heavy quark spin partners of the $T_{\bar c \bar c}(3875)^-$ and the $T_{cc}(3875)^+$.}
    \label{fig:TccPoleComparison}
\end{figure*}

By means of the approximation in Eq.~\eqref{eq:Sigapprox} for the $D^*D$ loop function embedded in the nuclear medium, and a similar one for that of the $\overline{D}{}^{\ast}\overline{D}$ meson pair, we can now
compute the isoscalar \texorpdfstring{$D^{\ast} D$}{D*D(*)} [$T(s\,;\,\rho)$] and  \texorpdfstring{$\overline{D}{}^{\ast}\overline{D}$}{barD*barD(*)} [$\overline{T}(s\,;\,\rho)$] scattering amplitudes inside of the nuclear environment in the whole complex
plane, for different medium densities $\rho$ and vacuum probabilities
$P_0$. We search for poles in the complex plane and find a pole on the first Riemann sheet (as defined in  Ref.~\cite{Albaladejo:2021cxj}) of the $T(s\,;\,\rho)$ and  $\overline{T}(s\,;\,\rho)$ amplitudes, off the real axis.\footnote{This does not
represent any violation of the analyticity properties of the
complete-system scattering $T$ matrices, because of the effective
procedure used to take into account the many body channels
of the type $D^*D N \to D^*D N'$ and $\overline{D}{}^{\ast}\overline{D} N \to \overline{D}{}^{\ast}\overline{D} N'$.} These complex poles are displayed in Fig.~\ref{fig:TccPoleComparison}, reinforcing the conclusions of the previous paragraph. A simple visual inspection of the two top plots of the figure clearly shows  the quite different $(\rho,P_0)$ pattern followed by the $T_{ c  c}^+$ and $T_{\bar c \bar c}^-$ poles produced by the presence of the nucleons. In general, the  $T_{ c  c}^+$ in the medium becomes broader than the $T_{\bar c \bar c}^-$, with the effective mass of the former (latter) displaced to higher (smaller) values than its nominal mass position in the free space. The future measurement of this  behavior should certainly shed light into the intricate dynamics of the $T_{ c  c}^+$ tetraquark-like state discovered by LHCb.

\subsection{\boldmath The \texorpdfstring{$T_{c c}^*(4016)^+$}{Tcc*+} and the \texorpdfstring{$T_{\bar c \bar c}^*(4016)^-$}{Tcc*-}}
\label{ss:resultsheavy-quark}

\begin{figure*}[ht]
    \centering
    \includegraphics[width=.45\textwidth]{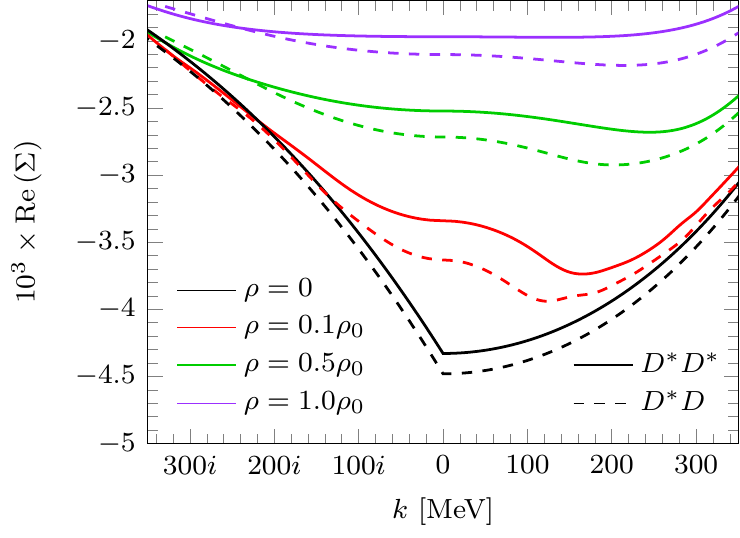}
    \hspace{5mm}
    \includegraphics[width=.45\textwidth]{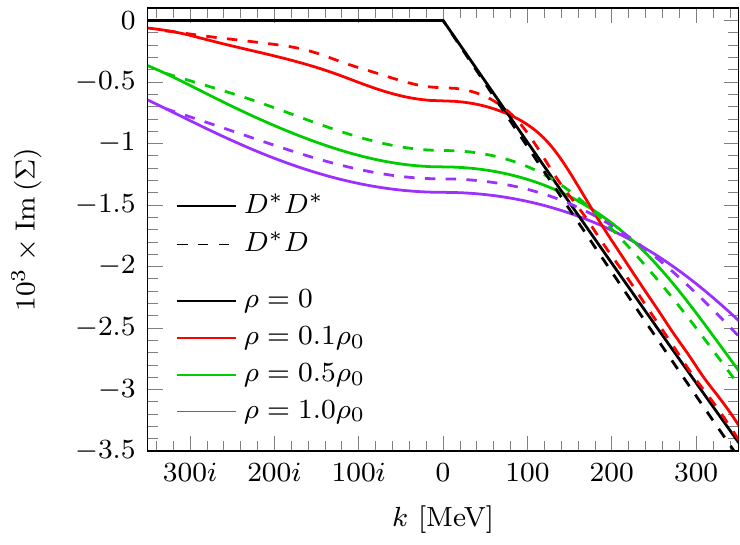}\\\vspace{0.5cm}
    \includegraphics[width=.45\textwidth]{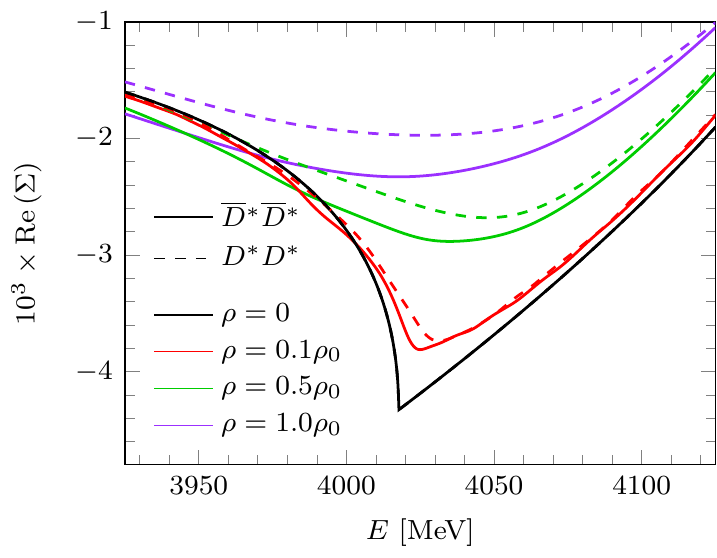}
    \hspace{5mm}
    \includegraphics[width=.45\textwidth]{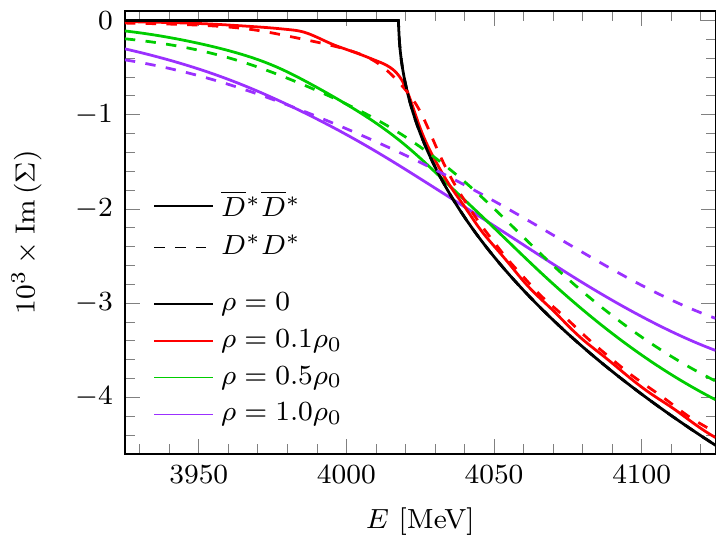}
    \caption{Top: Real (left) and imaginary (right) parts of the $D^* D^*$ (solid lines) and $D^*D$ (dashed lines) loop functions of Eqs.~\eqref{eq:DstarDstarLoopfunction} and \eqref{eq:DstarDLoopfunction}, respectively. We show results for different values of the nuclear medium density as a function of $k$, the c.m. three momentum of the heavy-light meson pair, since  the $D^* D^*$ and $D^*D$ thresholds are different.
    Bottom: Real (left) and imaginary (right) parts of the $\overline{D}{}^{\ast} \overline{D}{}^{\ast}$ (solid lines) and $D^*D^*$ (dashed lines) loop functions.  We show results for different values of the nuclear medium density as a function of the c.m. energy of the heavy-light meson pair.}
    \label{fig:starLoopComparison}
\end{figure*}

HQSS makes plausible the existence of an isoscalar $J^P=1^+$ $D^*D^*$ partner of the $T_{cc}(3875)^+$, which we have named as the  $T^*_{cc}(4016)^+$. It has been predicted by several theoretical groups~\cite{Albaladejo:2021vln,Du:2021zzh,Dai:2021vgf}, and  as discussed above in Subsect.~\ref{sec:hqss-part}, one should expect its mass to be higher than that of  the  $T_{ c  c}(3875)^+$ by an amount of the order $(m_{D^*}-m_D)\sim m_\pi$.\footnote{The situation here is necessarily similar to the open-charm sector, where one also notices $ m_{D_{s1}} - m_{D^{\ast}_{s0}} \simeq m_{D_1} - m_{D^{\ast}_0} \simeq m_{D^{\ast}_s} - m_{D_s} \simeq m_{D^{\ast}} - m_{D} \simeq m_\pi$ \cite{Albaladejo:2016lbb,Du:2017zvv,ParticleDataGroup:2022pth}.} In addition, the change of its properties inside of a nuclear medium will be also different to those described above for the   $T_{ c  c}^+$ since $D$ and $D^*$ spectral functions are different. From the comparison of  the top-left and bottom-left plots of Fig.~\ref{fig:TccPoleComparison} and the solid and dashed curves in the top plots of Fig.~\ref{fig:starLoopComparison}, we conclude that medium effects are larger for the  $T^*_{cc}(4016)^+$ than for the $T_{cc}(3875)^+$. This is because, within the model of Refs.~\cite{Garcia-Recio:2008rjt} and \cite{Romanets:2012hm}, the $D^*N\to D^*N$ interaction is stronger than the $DN\to DN$ one.

As it happened for the $T_{c c}(3875)^+$ and $T_{\bar c \bar c}(3875)^-$, the nuclear environment would induce different modifications to charmed $D^{*}D^*$ than to anti-charmed $\overline{D}{}^{\ast} \overline{D}{}^*$ pairs of interacting mesons, which will result into a different $(\rho,P_0)$ behavior for the $T^{*}_{\bar c \bar c}(4016)^-$, antiparticle of the $T^{*}_{cc}(4016)^+$, when it is produced in a nuclear medium.    This is now due to the different strength of the $D^{*}N$ and $\overline{D}{}^{*}N$ interactions. The bottom plots of  Figs.~\ref{fig:TccPoleComparison} and \ref{fig:starLoopComparison} illustrate the differences induced by the presence of nuclear matter, which become larger as the density and molecular probability increase. The nuclear medium breaks the particle-antiparticle symmetry leading to quite different $D^{(*)}$ and  $\overline{D}{}^{(*)}$ spectral functions.

%\newpage
\section{Conclusions} \label{sec:conclusions}

We have studied the behavior of the $T_{cc}(3875)^+$ and the $T_{\bar c\bar c}(3875)^-$ in the nuclear environment. We have considered both states to be isoscalar $S$-wave bound states that are generated as poles in the $D^{\ast} D$ and $\overline{D}{}^{\ast} \overline{D}$ scattering amplitudes, respectively. The in-medium effects have been incorporated by dressing the $D^{\ast} D$ and $\overline{D}{}^{\ast} \overline{D}$ loop functions with the corresponding spectral functions of the charmed mesons. We have then analyzed the $D^{\ast} D$ and $\overline{D}{}^{\ast} \overline{D}$ amplitudes in matter for energies around the common in-vacuum mass of the $T_{cc}(3875)^+$ and the $T_{\bar c\bar c}(3875)^-$ states so as to determine the modification of the pole positions in the medium.

For the interaction kernel we have considered two families of energy dependent interactions, consistent with heavy-quark spin symmetry, that allow for the analysis of the molecular probability content of these states. Indeed, the different analytical properties of these interactions manifest clearly at finite density, thus permitting to explore the connection between the in-medium behavior of the $T_{cc}(3875)^+$ and the $T_{\bar c\bar c}(3875)^-$ states and their nature. 

In contrast to low molecular probabilities, we have found that the medium effects on the $T_{cc}(3875)^+$ and the $T_{\bar c\bar c}(3875)^-$ amplitudes are sizable when large values of the molecular component are considered, leading to large widths for both states and shifts in mass at finite density with respect to their nominal values. In addition and due to the different nature of the $D^{(*)} N$ and $\overline{D}{}^{(\ast)} N$ interactions,  the $T_{cc}(3875)^+$ and $T_{\bar c\bar c}(3875)^-$ states behave differently in matter. By analysing the evolution with density of the  states in the complex energy plane we have seen very distinctive patterns. As a general rule, the $T_{cc}(3875)^+$ in matter becomes broader than the $T_{\bar c\bar c}(3875)^-$, whereas the mass of the former is moved to larger values than the nominal mass and the mass of the latter is displaced to smaller ones. Therefore, we expect that future measurements of these states in dense matter will give some important insights into their nature and their molecular content.

Finally, taking advantage of HQSS, we have also performed similar studies for the isoscalar $J^P=1^+$ HQSS partners of the $T_{cc}^+$ ($T_{cc}^{*+}$) and the $T_{\bar c\bar c}^-$ ($T_{\bar c\bar c}^{*-}$) by considering the $D^{\ast}D^{\ast}$ and $\overline{D}{}^{\ast}\overline{D}{}^{*}$ scattering amplitudes. We have found that the medium effects become larger for the $T^*_{cc}(4016)^+$ than for the $T_{cc}(3875)^+$, as the $D^*N \rightarrow D^*N$ interaction is stronger than the $DN \rightarrow DN$ one. Also, similarly to the $T_{cc}(3875)^+$ and $T_{\bar c\bar c}(3875)^-$ states, the different strength of the $D^{\ast}N$ and $\overline{D}{}^{\ast}N$ interactions leads to a distinctive behavior of the $T^{\ast}_{cc}(4016)^+$ and its antiparticle with density, specially for large values of the molecular content. 

All in all, we can conclude that an interesting venue to discern the molecular nature of $T_{cc}(3875)^+$, the $T_{\bar c\bar c}(3875)^-$, and their HQSS partners would be to experimentally determine their behavior in a dense nuclear environment, such as the one generated in HICs under the expected conditions at the CBM (FAIR) or with fixed nuclear targets such as  $\bar p$-nuclei in PANDA (FAIR).

\acknowledgements
This work was supported by the Spanish Ministerio de Ciencia e Innovaci\'on (MICINN) under contracts No.\, PID2019-110165GB-I00 and No.\, PID2020-112777GB-I00, by Generalitat Valenciana under contract PROMETEO/2020/023, and from the project CEX2020-001058-M Unidad de Excelencia ``Mar\'{\i}a de Maeztu"). This project has received funding from the European Union Horizon 2020 research and innovation programme under the program H2020-INFRAIA-2018-1, grant agreement No.\,824093 of the STRONG-2020 project.  M.\,A. and V.\,M.~are supported through Generalitat Valenciana (GVA) Grants No.\,CIDEGENT/2020/002 and ACIF/2021/290, respectively. %and thanks the warm support of ACVJLI. 
L.\,T. also acknowledges support from the CRC-TR 211 'Strong-interaction matter under extreme conditions'- project Nr. 315477589 - TRR 211 and from the Generalitat de Catalunya under contract 2021 SGR 171.

\bibliographystyle{JHEP}
\bibliography{references.bib}

\end{document}